\documentclass[aip,reprint]{revtex4-1}

\usepackage{graphicx}
\usepackage{dcolumn}
\usepackage{bm}

\usepackage[utf8]{inputenc}
\usepackage[T1]{fontenc}
\usepackage{mathptmx}
\usepackage{etoolbox}

\usepackage{xcolor}

\usepackage{amsmath}

\DeclareMathOperator{\erf}{erf}

\newcommand{\asin}{\text{arcsin}}

\usepackage{hyperref}
\hypersetup{
        colorlinks=true,       
        linkcolor=blue,          
        citecolor=blue,        
        filecolor=magenta,      
        urlcolor=blue
        }

\begin{document}

\title{Electron Beam Characterization via Quantum Coherent Optical Magnetometry}

\author{Nicolas DeStefano}
\email{ncdestefano@wm.edu}
\author{Saeed Pegahan}
\affiliation{Dept. of Physics, William \& Mary, Williamsburg, Virginia 23187, USA}
\author{Aneesh Ramaswamy}
\affiliation{Stevens Institute of Technology, Hoboken, New Jersey 07030, USA}
\author{Seth Aubin}
\author{T. Averett}
\affiliation{Dept. of Physics, William \& Mary, Williamsburg, Virginia 23187, USA}
\author{Alexandre Camsonne}
\affiliation{Thomas Jefferson National Accelerator Facility, Newport News, Virginia 23606, USA}
\author{Svetlana Malinovskaya}
\affiliation{Stevens Institute of Technology, Hoboken, New Jersey 07030, USA}
\author{Eugeniy E. Mikhailov}
\affiliation{Dept. of Physics, William \& Mary, Williamsburg, Virginia 23187, USA}
\author{Gunn Park}
\author{Shukui Zhang}
\affiliation{Thomas Jefferson National Accelerator Facility, Newport News, Virginia 23606, USA}
\author{Irina Novikova}
\affiliation{Dept. of Physics, William \& Mary, Williamsburg, Virginia 23187, USA}

\date{\today}

\begin{abstract}
We present a quantum optics-based detection method for determining the position and current of an electron beam. As electrons pass through a dilute vapor of rubidium atoms, their magnetic field perturb the atomic spin’s quantum state and causes polarization rotation of a laser resonant with an optical transition of the atoms. By measuring the polarization rotation angle across the laser beam, we recreate a 2D projection of the magnetic field and use it to determine the e-beam position, size and total current. We tested this method for an e-beam with currents ranging from 30 to 110~$\mu$A. Our approach is insensitive to electron kinetic energy, and we confirmed that experimentally between 10 to 20~keV. This technique offers a unique platform for non-invasive characterization of charged particle beams used in accelerators for particle and nuclear physics research. 

\end{abstract}

\maketitle

With the advent of charged particle accelerators came the need for accurate beam diagnostics. Driven by improvements to quality and control of charged particle beams, the sensitivity and precision of in-situ diagnostics must meet the needs of new particle accelerators where increasingly strict demands are placed on beam properties such as energy, current, emittance and others. The need for increasingly precise beam diagnostics for a wide range of parameters inspires relentless efforts to continue towards the  development of more robust, non-invasive spatial beam parameter measurements. Much of this development is focused on the methods relying on optical signals. Synchrotron radiation has been used for monitoring beam position and size \cite{Bossart1980, Wang2013, WLi2022}, and laserwires~\cite{Hofmann2018, Corner2014} rely on Compton scattering from a high-power laser to extract the charged particle beam parameters. However, these types of optical diagnostics are limited by technical requirements on the particle and laser beams. For example, synchrotron radiation is only available in particle trajectory-bending components such as magnets, while the laserwire method requires a high-intensity laser to slowly scan the beam profile, and often requires additional radiation, or electronic, detectors. Beam profile monitors based on gas ionization and excitation by particle beams have been demonstrated and used at different accelerators \cite{Anne1993, Sandberg2021, Variola2007}. More recently, a 2D beam monitor measuring fluorescence from the interaction between a 5 keV particle beam and a supersonic gas curtain was reported \cite{Salehilashkajani2022}, but it is not suitable for spatial longitudinal profile measurement, and the sensitivity is quite limited. It is worth mentioning another type of device widely used in the accelerator community, the RF-based beam monitor, which can provide beam centroid at high resolution, but is incapable of profile measurement \cite{Maesaka2012}. 

\begin{figure*}
    \centering
    \includegraphics[width=0.85\textwidth]{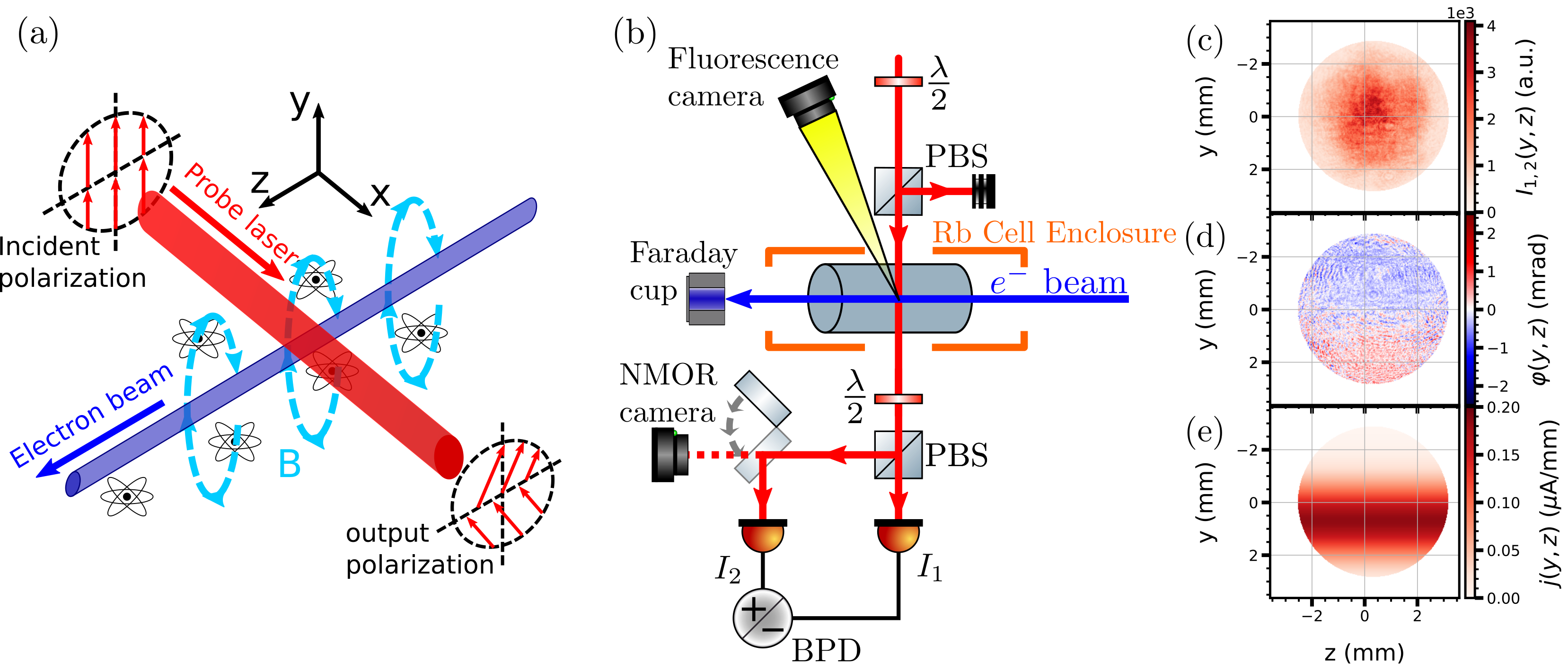}
    \caption{(a) The basic concept of the charged particle beam detection method. A linear polarization of a laser beam (red) is affected by the magnetic field (dashed light blue circles) of an electron beam (dark blue) mediated by the spin coherence of Rb atoms. (b) Schematic of the experimental setup (see text for abbreviations), where a flipper mirror determines BPD- or CCD camera-based detection. (c) Laser intensity profile at the output of the PBS, recorded by the CCD with $\sim 200\  \mu$s exposure time. (d) The $e$-beam-induced polarization rotation angle, $\varphi(y,z)$, calculated using  Eq.~\ref{eqn:rotation_equation_intensities}. (e) The electron current density distribution reconstructed from the $\erf$-function fit of the normalized polarization rotation $\Phi(y,z)$.  For all the image analysis an intensity mask is applied to eliminate data points with laser intensity below $5$\% of the peak value to prevent infinities arising from Eq.~\ref{eqn:rotation_equation_intensities}}
    \label{fig:experiment_summary}
\end{figure*}

In this paper, we propose a qualitatively different approach to beam diagnostics that takes advantage of recent advances in quantum atom-based optical sensors to map the magnetic field produced by the moving charged particles and then reconstruct the beam parameters. In this proof-of-principle demonstration, we use coupling between resonant laser light and atomic spins to monitor  evolution of the latter in the magnetic field of a collimated electron beam. The essence of the proposed approach is shown in Fig.~\ref{fig:experiment_summary}(a). The $e$-beam travels through a cell containing a dilute gas of rubidium  atoms. Within the cell each Rb spin precesses at a rate determined by the local magnetic field. A linearly polarized laser beam traverses the volume surrounding the charge particle beam and probes the atomic spins: the nonlinear magneto-optical polarization rotation (NMOR) effect~\cite{Budker2013,Weis2017,Fu2020} rotates the optical linear polarization axis due to the magnetic field of the $e$-beam. By measuring the polarization rotation variation across the laser beam cross-section, we are able to determine the local magnetic field of the $e$-beam and reconstruct its transverse spatial profile. Our detection scheme is largely non-invasive as the $e$-beam is minimally affected by low-density alkali vapor ($10^{10}-10^{12}~\mathrm{cm}^{-3}$) localized within a 30 cm region enclosed in high-vacuum\footnote{We calculate a Bethe energy loss of 0.1 eV/m at a Rb density of $10^{12}~\mathrm{atoms/cm}^3$ for 20 keV electrons. At 1 GeV, the energy loss is 0.04 eV/m.}. The strong resonant coupling between laser light and atomic spin coherence enhances the sensitivity compared to the approaches based on incoherent electron impact-induced fluorescence \cite{Salehilashkajani2022}. This demonstration is a stepping stone toward more sensitive and comprehensive detection of charged particle beams using advanced spectroscopy techniques.

The proposed $e$-beam detector relies on two effects: high sensitivity of atomic spin state superposition to the magnetic field and strong dependence of atoms' resonant optical properties on their spin state. Thanks to the Zeeman effect,  the energy sub-levels with different magnetic quantum numbers, $m$, shift by different amounts, a superposition of two such sub-levels evolves in time, developing a magnetic field dependent relative phase. A resonant, and linearly polarized laser field, can simultaneously prepare the desired quantum superposition and measure its evolution. Indeed, the two circularly-polarized components create a two-photon transition between the states with $m = \pm 1$ (spin alignment, see Fig. S1 in Supplementary). The optical field's propagation through the magnetic field changes the relative phase between the two circular polarization components, resulting in rotation of the original linear polarization. This effect, known as nonlinear magneto-optical polarization rotation (NMOR), is a convenient and sensitive method for optical magnetic field measurements. In the case of long spin coherence lifetime and exact optical resonance, the rate at which the polarization rotation angle $\varphi$ rotates (as the laser propagates along the $x$-axis) is proportional to the local magnetic field $B$ (see Supplementary A for derivation):

\begin{equation}
\label{eq:polarizationRotation_equation}
\left. \frac{d\varphi}{dx}\right|_{B\approx 0}=\frac{\hbar c N}{\lambda I} \gamma B,
\end{equation}

where we assume the greatest contribution to $\varphi$ is when the probe beam direction and magnetic field $B$ are along the $x$-axis; $\lambda$, $I$, and $c$ are the wavelength, local intensity, and speed of light, respectively; $N$ is the density of the Rb vapor, and $\gamma=5$~Hz/nT is the gyromagnetic ratio for $^{85}$Rb atoms. In principle, since the intrinsic spin coherence lifetime is very long, and in some experiments \cite{BalabasPRL2010} was extended up to many seconds, NMOR-based magnetometers can achieve an impressive sub-pT sensitivity~\cite{Rosner2022,Lucivero2021}. When the spin decoherence and optical losses are accounted for, the exact proportionality coefficient between the polarization rotation rate and the applied magnetic field is more complex and depends, at some extent, on many experimental parameters (lifetimes of both optical spin states, laser frequency detuning from the optical resonance, influence of additional near-resonant atomic levels, etc. See Supplementary A for detailed derivations). We experimentally derive the rotation response profile $\beta(y,z) =  \frac{1}{B}\frac{d\varphi}{dx}$ across the Gaussian laser beam by applying a known constant magnetic field and measure the polarization rotation for each camera pixel. 

In the experiment, the laser beam propagates along the $x$-axis shown in Fig.~\ref{fig:experiment_summary}(a), nearly perpendicular to the electron beam direction, designated as the $z$-direction. The laser is linearly polarized in the $y$-direction, perpendicular to both  the electron and the laser beam propagation directions.  In this configuration, the laser polarization rotation should only be sensitive to the longitudinal magnetic field component $B_x$ (Faraday configuration)~\cite{Budker_RMP2002}. For a narrow laser beam the magnitude and sign of the rotation angle depend on the cumulative magnetic field along the optical propagation path. Thus, imaging a 2D map of the polarization angle on a camera using a laser beam with large cross-section, we can in principle obtain a transverse profile of the $e$-beam. The schematic of the experimental setup is shown in Fig.~\ref{fig:experiment_summary}(b). 

We used a commercial thermionic electron source generating a collimated electron beam with energy 10-20 keV and current up to 200~$\mu$A. The electron beam passes through a glass cell (inner dimensions: 10 mm $\times$ 10 mm $\times$ 45 mm) containing rubidium vapor before terminating in a Faraday cup.  Differential pumping through 8 mm apertures, up- and down-stream of the cell, are used to confine the vapor to the cell and to keep it at constant pressure. The vapor was maintained at $\sim 60^\circ$C, corresponding to a $^{85}$Rb vapor density of $2.7\times10^{11}\mathrm{cm}^{-3}$ and a pressure of $10^{-6}-10^{-5}$~Torr, housed within a vacuum system of pressure $10-20$ nTorr during experiment operation to preserve the quality of the electron source filament. Since such system is most sensitive to small magnetic fields, so we suppress environmental fields by placing the atomic vapor inside a layer of $\mu$-metal magnetic shielding, and use external coils to further reduce the background magnetic fields. 

For optical detection, we use an external cavity diode laser (ECDL) operating at the $D_2$ line of ${}^{85}$Rb (wavelength $\lambda = 780$~nm), specifically at the $5^2 S_{1/2}, F=3 \to 5^2 P_{3/2}, F'$ transition. The laser beam was linearly polarized using a polarizing beam splitter (PBS) cube and enlarged to 6~mm before the entering the Rb-filled glass cell to capture the full $e$-beam diameter. Beyond the cell, the laser polarization rotation is analyzed using a balanced polarimeter, consisting of a half-wave plate that rotates the polarization by $45^\circ$, an analyzer PBS and a differential amplified photodetector (BPD). In the absence of the $e$-beam, the intensities of the two PBS outputs $I_{1,2}$ are balanced. A small rotation of the polarization $\varphi$ produces a proportional variation between the two channel intensities, allowing accurate calculations of $\varphi$:

\begin{equation}
    \varphi = \asin \left( \frac{I_2 - I_1}{2 (I_2 + I_1)} \right) \approx \frac{I_2 - I_1}{2 (I_2 + I_1)}.
    \label{eqn:rotation_equation_intensities}
\end{equation}

In our experiment we measure total power changes in each channel that let us measure the integrated rotation signal, which is convenient for system alignment. For most presented data we use a CCD-based imaging system (magnification 0.50) to record the spatial distribution $\varphi(y,z)$ across the laser beam. To ensure the consistency between the recorded intensity masks $I_{1,2}$ they are recorded consecutively at the same camera position, but with a different angle of the waveplate before the polarizer. An example of such intensity profile for one of the channels is shown in Fig.~\ref{fig:experiment_summary}(c), in which the intensity difference between two channels, induced by the $e$-beam, is not distinguishable by the naked eye. As the laser beam has Gaussian intensity profile, we only use its central part with sufficient intensity in further calculations. In addition, the electron beam was pulsed on and off at $1$Hz to record two consecutive images, with and without the $e$-beam. Subtracting the images from each other allows us to remove any residual polarization rotation due to stray magnetic fields or optical elements to ensure that we only detect the polarization rotation caused by the $e$-beam. 

By capturing the intensity profiles of the two outputs on a CCD camera with two waveplate positions, we were able to calculate local variations of $\varphi(y,z)$ within the laser beam cross-section as shown in Fig.~\ref{fig:experiment_summary}(d). The angle distribution within the laser beam is extracted by applying Eq.~\ref{eqn:rotation_equation_intensities} to each camera pixel. Since the $e$-beam generates a circulating magnetic field in $x-y$ plane, and the nonlinear polarization rotation is primarily sensitive to the $x$ component of the magnetic field, we expect to see a sign change in the measured polarization rotation angle below and above the center of the $e$-beam,  as shown in Fig.~\ref{fig:experiment_summary}(a). 

\begin{figure*}
    \centering
    \includegraphics[width=0.9\textwidth]{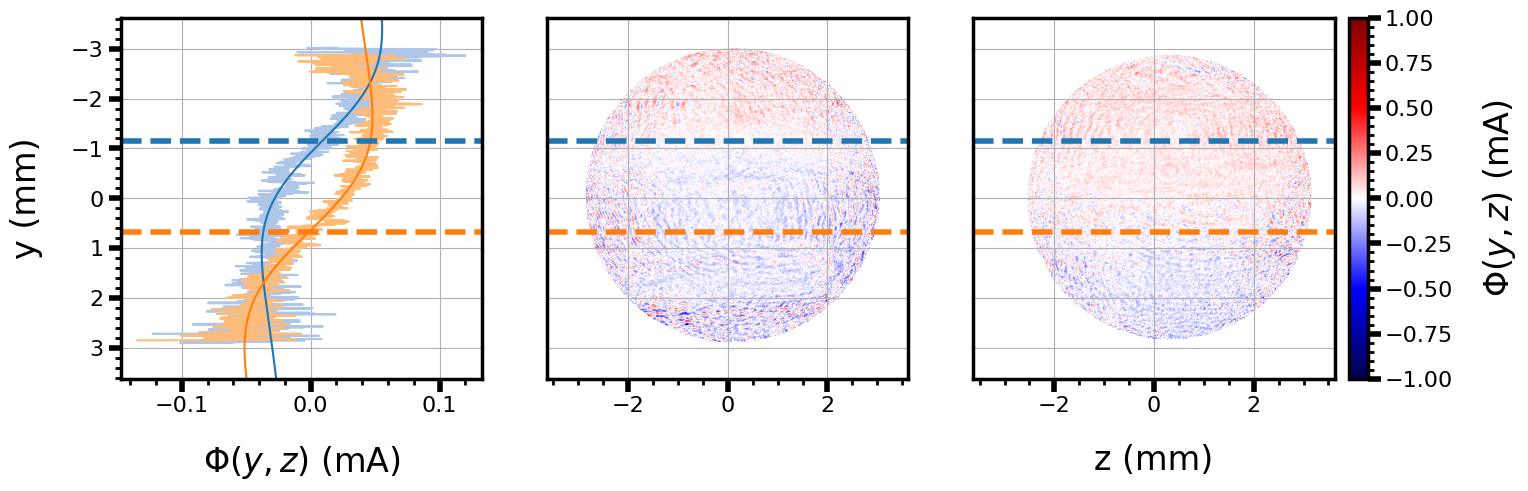}
    \caption{Measured normalized rotation images $\Phi(y,z)$ for two different electron beam positions of electron beam emission current $\mathcal{I}_E=200\mu$A and energy $E = 20$ keV. The location of the $e$-beam center in each case is clearly detectable by the reversal of the polarization rotation direction. The left panel shows the vertical profiles of the images with the corresponding fits from Eq. \ref{eq:PolRot}, and the horizontal dashed lines indicate the positions of the electron beam center, also extracted from the fits.}
    \label{fig:imaging_results}
\end{figure*}

To obtain more quantitative information about the $e$-beam, we assume that its current density $j(x,y)$ is cylindrically symmetric and has a Gaussian transverse profile: 

\begin{equation}
    j(x,y)=\frac{\mathcal{I}_0}{\pi w^2}e^{-\frac{x^2+y^2}{w^2}},
\end{equation}\label{eq:GsCurrentDens}

where $\mathcal{I}_0$ is the total current, and $w$ is the beam $1/e^2$ half-width. The corresponding magnetic field maintains cylindrical symmetry, forming concentric field lines around the beam central axis, and its magnitude can be easily found using Ampere's law:

\begin{equation}
    B(x,y)=\frac{\mu_0\mathcal{I}_0}{2 \pi \sqrt{x^2+y^2}}\left(1-e^{-\frac{x^2+y^2}{w^2}}\right), \label{eq:GsMagnField}
\end{equation}

where $\mu_0$ is the permeability of free space.

In the limit of a weak magnetic field, the total measured polarization rotation is integrated along the laser probe propagation path $L$:

\begin{equation}
    \varphi(y,z)=\beta(y,z) \int_{-L/2}^{L/2} B_x(x,y)dx.
\end{equation}

Here we assume that the rotation angle is small, the $e$-beam is collimated and its magnetic field has no $z$-dependence. In this case any changes in the polarization rotation in this direction can only be caused by the variation in the atomic response $\beta(y,z)$ due to, e.g., laser intensity variation. Using Eq.~\ref{eq:GsMagnField}, and assuming that the length of the cell is much larger than the $e$-beam width $L\gg w$, we can find an analytical expression for the polarization rotation for an electron beam centered at vertical location $y=y_0$:

\begin{eqnarray}
  &&  \varphi(y,z)=\beta(y,z) \frac{\mu_0\mathcal{I}_0(y-y_0)}{2 \pi}\int_{-L/2}^{L/2} \frac{1-e^{-\frac{x^2+(y-y_0)^2}{w^2}}}{x^2+(y-y_0)^2}dx \nonumber \\ 
    \approx 
  &&  \frac{\beta\mu_0\mathcal{I}_0}{2} \left[\erf\left(\frac{y-y_0}{w}\right)-\frac{2}{\pi}\arctan\left(\frac{2(y-y_0)}{L}\right)\right], \label{eq:PolRot}
\end{eqnarray}

where $\erf(x)$ is the error function, and we assume $e^{-L^2/4w^2}\ll 1$. For a longer cell the first term dominates the rotation, while the edge effects become more noticeable farther from the $e$-beam center.

It is convenient to introduce a normalized signal $\Phi(y,z)=\varphi(y,z)/(\mu_0 \beta(y,z))$ since it depends only on the $e$-beam current distribution. For a Gaussian current distribution,  according to Eq.~\ref{eq:PolRot}, the normalized signal is an error function, centered at the vertical position of the $e$-beam $y_0$, and the maximum variation of $j(y,z)$ depends only on the total $e$-beam current, that allows for robust measurements of these parameters even for a noisy signals. An example of the electron current distribution using the parameters obtained from fitting the normalized rotation spectra are shown in  Fig.~\ref{fig:experiment_summary}(e).

\begin{figure}[t]
    \centering
    \includegraphics[width=0.9\linewidth]{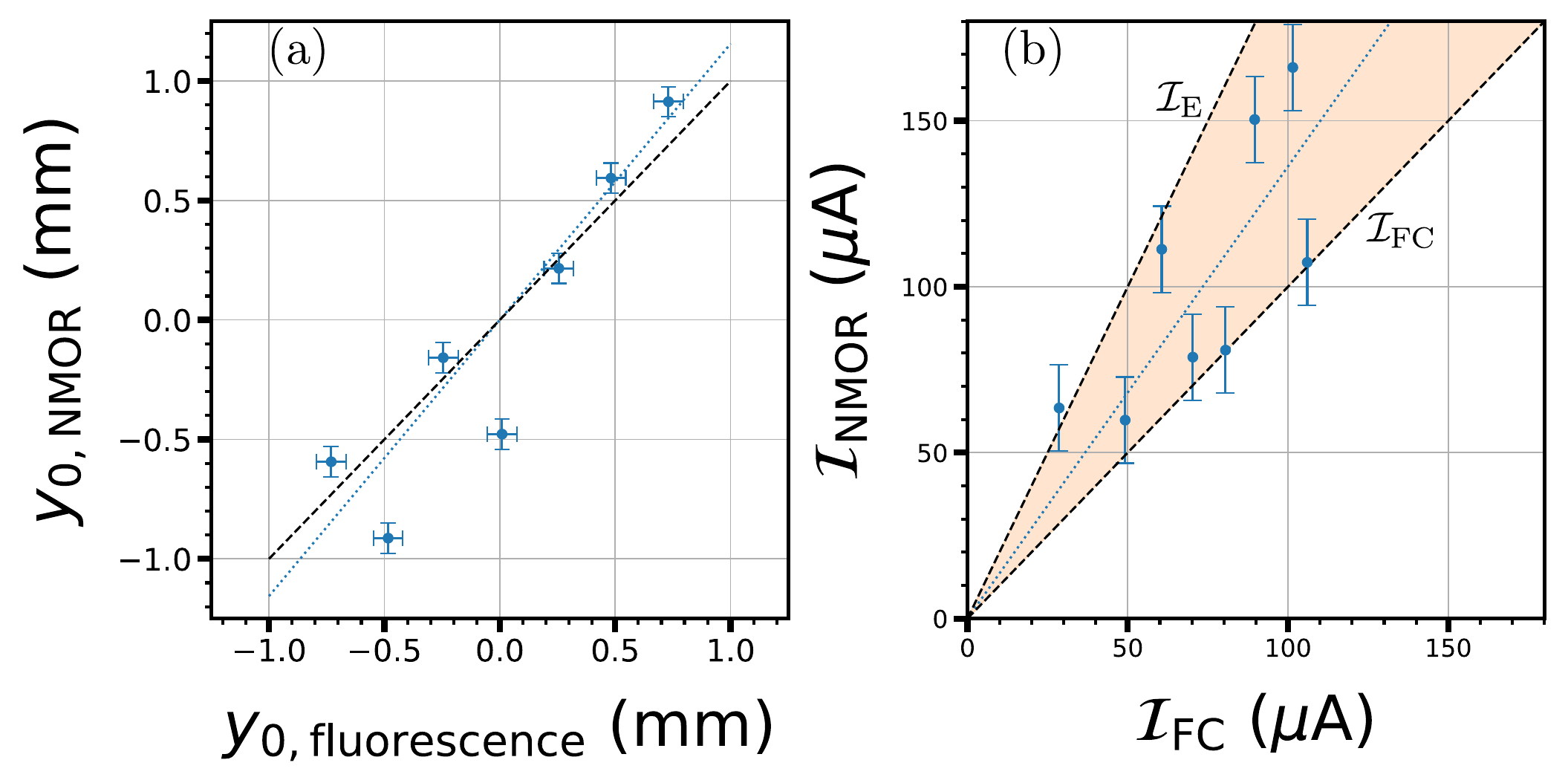}
    \caption{(a) Comparison between the electron beam center position extracted from the polarization rotation measurement $y_{0,\text{NMOR}}$ and from the electron-induced rubidium fluorescence images $y_{0,\text{fluorescence}}$ with a regression of $y_{0, \text{NMOR}} = (1.16 \pm 0.21)\ y_{0, \text{fluorescence}}$~(dotted blue line). (b) Comparison between the total electron beam current calculated using the polarization rotation fits $\mathcal{I}_{\text{NMOR}}$ and measured directly using the Faraday cup $\mathcal{I}_{\text{FC}}$ with a regression of $\mathcal{I}_{\text{NMOR}} = (1.36 \pm 0.13)\ \mathcal{I}_{\text{FC}}$~ (dotted blue line). The black dashed lines have a slope of 1 and 2, representing $\mathcal{I}_{\text{FC}}$ and the emission current $\mathcal{I}_{\text{E}} \approx 2 \mathcal{I}_{\text{FC}}$. The shaded region indicates the range of valid beam currents measurable by NMOR. Uncertainties in NMOR-derived parameters stem from the variance of electron beam center position and total current under identical experimental conditions. The fluorescence uncertainty is based on the variance in center position for varied beam currents, while the Faraday cup signal uncertainty is due to the variance of the measured Faraday cup signal (read on an oscilloscope).} 
    \label{fig:parameter_verification}
\end{figure}

Fig.~\ref{fig:imaging_results} shows the examples of the recorded normalized signal  $\Phi(y,z)$ for two positions of the $e$-beam. As expected, the polarization rotation changes direction from positive to negative at the $e$-beam center position, and the signal is uniform in the $z$-direction. To obtain the $e$-beam parameters we fit the 2D experimental signal distribution with the error function. We then repeated the measurements for varied electron beam positions and values of the total current, as shown in Fig.~\ref{fig:parameter_verification}. We verify the accuracy of the beam position measurements by capturing images of fluorescence from Rb atoms ionized by the $e$-beam (see Supplementary C for details). While both signals are noisy, the measured centroid position variation matches within 16\% between the two methods, indicating that the spatial coordinate systems agree between the two separate imaging systems. Similarly, we compare the total $e$-beam current value extracted from the polarization rotation measurements with that measured at the Faraday cup $\mathcal{I}_{\text{FC}}$ and electron emission current $\mathcal{I}_{\text{E}}$, located approximately 25~cm downstream of the Rb cell, matching within 36\% of $\mathcal{I}_{\text{FC}}$ and 14\% of $\mathcal{I}_{\text{E}}$ between the two methods. Varying the electron beam energy between 10 to 20 keV yielded no significant changes in the profiles or quantities derived using the polarization rotation method.

From the fit, we obtain a FWHM $e$-beam diameter of $1.96 \pm 0.13$ mm. Although we are unable to independently verify the precise profiles of the electron beam, this value is noticeably broader than that obtained from the fluorescence fits, $0.89 \pm 0.04$ mm FWHM. The uncertainty for these quantities are derived by considering the variance of the profile parameters of the datasets. We attribute this discrepancy between the widths in part to poor signal-to-noise ratio (SNR) of the rotation signal, especially at the edges of the laser beam, where the laser intensity is low. The overall rotation signal was typically below 1~mrad, and thus strongly affected by the camera electronic noise. Further, the presence of a transverse magnetic field $B_y$ has been shown to broaden the NMOR resonance \cite{Anisimov2003}, and thus the unaccounted transverse components of the magnetic field can increase the estimated width of the electron beam. In the future we plan  improve the accuracy of the electron beam width measurements by, for example, realizing a more complex optical interrogation scheme to enhance atomic spin coherence and boost the magnetic response of atoms~\cite{Zhuarxiv1804.08194,Zhu:21}, as well as by developing a model describing the polarization rotation for arbitrary magnetic field orientation. Also, using a low-noise CCD camera and pulsing the $e$-beam at higher rate will remove the dominant source of the technical noise and may boost the sensitivity by several order of magnitude, limited by the laser shot noise (see Supplementary B for details). Moreover, in this case, we may be able further improve the performance by using a non-classical (squeezed) optical field~\cite{Wolfgramm2010,Horrom2012,OtterstromOL2014,BaiJOpt2021,Li2022,WuSA2023}. The ultimate spatial resolution of our method is diffraction limited and can potentially resolve details down to a few microns. Using these enhancements, we can accurately image the current distribution of the electron beam non-invasively with optimum SNR. 

Looking forward, higher sensitivity and spatial resolution could be achieved by employing advanced spectroscopic methods based on two or more lasers. For example, the 2D transverse current distribution can potentially be mapped out by 4-wave mixing, in which two intersecting probe lasers generate a third beam (imaged on a camera) that depends on the magnetic field in the crossing region~\cite{BoyerLett_PRL2007}. Currently we are investigating 2- and 3-photon excitation to Rydberg states where we expect much higher sensitivity to the magnetic and electric fields from the $e$-beam due to their large Stark and Zeeman shifts~\cite{Raithel_PRA2019, Haroche_NatPhys2019}. Indeed, Rydberg states of ultracold atoms may have sufficient sensitivity for single particle detection~\cite{Raizen_NJP2010,Kawasaki_PhysRevRes2023}. 

In summary, thanks to nonlinear interaction with atomic spins, the light polarization rotates in a dilute alkali-metal vapor in the presence of the magnetic field produced by a passing charged particle beam. Taking advantage of this effect, we demonstrated a non-invasive method for characterizing the position and total current of an electron beam that was obtained by mapping the nonlinear polarization rotation of a transverse probe laser crossing the $e$-beam. We experimentally evaluated the accuracy of the proposed technique to determine the total current and transverse position at 20 keV for $e$-beam currents between 30--110 $\mu$A, and discussed its current limitations.  Since the $e$-beam is detected via its magnetic field, the scheme is insensitive to the beam energy \cite{Ramaswamy2022}, charged particle type and local electric fields associated with the beam. We expect that this technique can be applied to high energy particle accelerators and also refined to meet the precision required for experiments at the frontier of nuclear and high-energy physics research.

\section*{Supplementary Material}
See the supplementary material for detailed discussion of NMOR, sensitivity, and electron-induced rubidium fluorescence.

\begin{acknowledgments}
This work is supported by U.S. DOE Contract No. DE-AC05-06OR23177, NSF award 2326736 and Jefferson Lab LDRD program. The collaboration thanks Jiahui Li and Cutter Fugett for their assistance in the early stages of the project.
\end{acknowledgments}

\subsection*{Data Availability Statement}
The data that support the findings of this study are available from the corresponding author upon reasonable request.

\section*{References}
\bibliography{bibliography}

\begin{thebibliography}{33}%
\makeatletter
\providecommand \@ifxundefined [1]{%
 \@ifx{#1\undefined}
}%
\providecommand \@ifnum [1]{%
 \ifnum #1\expandafter \@firstoftwo
 \else \expandafter \@secondoftwo
 \fi
}%
\providecommand \@ifx [1]{%
 \ifx #1\expandafter \@firstoftwo
 \else \expandafter \@secondoftwo
 \fi
}%
\providecommand \natexlab [1]{#1}%
\providecommand \enquote  [1]{``#1''}%
\providecommand \bibnamefont  [1]{#1}%
\providecommand \bibfnamefont [1]{#1}%
\providecommand \citenamefont [1]{#1}%
\providecommand \href@noop [0]{\@secondoftwo}%
\providecommand \href [0]{\begingroup \@sanitize@url \@href}%
\providecommand \@href[1]{\@@startlink{#1}\@@href}%
\providecommand \@@href[1]{\endgroup#1\@@endlink}%
\providecommand \@sanitize@url [0]{\catcode `\\12\catcode `\$12\catcode
  `\&12\catcode `\#12\catcode `\^12\catcode `\_12\catcode `\%12\relax}%
\providecommand \@@startlink[1]{}%
\providecommand \@@endlink[0]{}%
\providecommand \url  [0]{\begingroup\@sanitize@url \@url }%
\providecommand \@url [1]{\endgroup\@href {#1}{\urlprefix }}%
\providecommand \urlprefix  [0]{URL }%
\providecommand \Eprint [0]{\href }%
\providecommand \doibase [0]{http://dx.doi.org/}%
\providecommand \selectlanguage [0]{\@gobble}%
\providecommand \bibinfo  [0]{\@secondoftwo}%
\providecommand \bibfield  [0]{\@secondoftwo}%
\providecommand \translation [1]{[#1]}%
\providecommand \BibitemOpen [0]{}%
\providecommand \bibitemStop [0]{}%
\providecommand \bibitemNoStop [0]{.\EOS\space}%
\providecommand \EOS [0]{\spacefactor3000\relax}%
\providecommand \BibitemShut  [1]{\csname bibitem#1\endcsname}%
\let\auto@bib@innerbib\@empty
\bibitem [{\citenamefont {Bossart}\ \emph {et~al.}(1980)\citenamefont
  {Bossart}, \citenamefont {Bosser}, \citenamefont {Burnod}, \citenamefont
  {d'Amico}, \citenamefont {Ferioli}, \citenamefont {Mann}, \citenamefont
  {Meot},\ and\ \citenamefont {Co{\"i}sson}}]{Bossart1980}%
  \BibitemOpen
  \bibfield  {author} {\bibinfo {author} {\bibfnamefont {R.}~\bibnamefont
  {Bossart}}, \bibinfo {author} {\bibfnamefont {J.}~\bibnamefont {Bosser}},
  \bibinfo {author} {\bibfnamefont {L.}~\bibnamefont {Burnod}}, \bibinfo
  {author} {\bibfnamefont {E.}~\bibnamefont {d'Amico}}, \bibinfo {author}
  {\bibfnamefont {G.}~\bibnamefont {Ferioli}}, \bibinfo {author} {\bibfnamefont
  {J.}~\bibnamefont {Mann}}, \bibinfo {author} {\bibfnamefont {F.}~\bibnamefont
  {Meot}}, \ and\ \bibinfo {author} {\bibfnamefont {R.}~\bibnamefont
  {Co{\"i}sson}},\ }\enquote {\bibinfo {title} {Proton beam profile monitor
  using synchrotron light},}\ in\ \href {\doibase 10.1007/978-3-0348-5540-2_60}
  {\emph {\bibinfo {booktitle} {11th International Conference on High-Energy
  Accelerators: Geneva, Switzerland, July 7--11, 1980}}},\ \bibinfo {editor}
  {edited by\ \bibinfo {editor} {\bibfnamefont {W.~S.}\ \bibnamefont
  {Newman}}}\ (\bibinfo  {publisher} {Birkh{\"a}user Basel},\ \bibinfo
  {address} {Basel},\ \bibinfo {year} {1980})\ pp.\ \bibinfo {pages}
  {470--475}\BibitemShut {NoStop}%
\bibitem [{\citenamefont {Wang}\ \emph {et~al.}(2013)\citenamefont {Wang},
  \citenamefont {Rubin}, \citenamefont {Conway}, \citenamefont {Palmer},
  \citenamefont {Hartill}, \citenamefont {Campbell},\ and\ \citenamefont
  {Holtzapple}}]{Wang2013}%
  \BibitemOpen
  \bibfield  {author} {\bibinfo {author} {\bibfnamefont {S.}~\bibnamefont
  {Wang}}, \bibinfo {author} {\bibfnamefont {D.}~\bibnamefont {Rubin}},
  \bibinfo {author} {\bibfnamefont {J.}~\bibnamefont {Conway}}, \bibinfo
  {author} {\bibfnamefont {M.}~\bibnamefont {Palmer}}, \bibinfo {author}
  {\bibfnamefont {D.}~\bibnamefont {Hartill}}, \bibinfo {author} {\bibfnamefont
  {R.}~\bibnamefont {Campbell}}, \ and\ \bibinfo {author} {\bibfnamefont
  {R.}~\bibnamefont {Holtzapple}},\ }\bibfield  {title} {\enquote {\bibinfo
  {title} {Visible-light beam size monitors using synchrotron radiation at
  cesr},}\ }\href {\doibase https://doi.org/10.1016/j.nima.2012.11.097}
  {\bibfield  {journal} {\bibinfo  {journal} {Nuclear Instruments and Methods
  in Physics Research Section A: Accelerators, Spectrometers, Detectors and
  Associated Equipment}\ }\textbf {\bibinfo {volume} {703}},\ \bibinfo {pages}
  {80--90} (\bibinfo {year} {2013})}\BibitemShut {NoStop}%
\bibitem [{\citenamefont {Li}\ \emph {et~al.}(2022)\citenamefont {Li},
  \citenamefont {Yan}, \citenamefont {Liu},\ and\ \citenamefont
  {Wu}}]{WLi2022}%
  \BibitemOpen
  \bibfield  {author} {\bibinfo {author} {\bibfnamefont {W.}~\bibnamefont
  {Li}}, \bibinfo {author} {\bibfnamefont {J.}~\bibnamefont {Yan}}, \bibinfo
  {author} {\bibfnamefont {P.}~\bibnamefont {Liu}}, \ and\ \bibinfo {author}
  {\bibfnamefont {Y.~K.}\ \bibnamefont {Wu}},\ }\bibfield  {title} {\enquote
  {\bibinfo {title} {Synchrotron radiation interferometry for beam size
  measurement at low current and in large dynamic range},}\ }\href {\doibase
  10.1103/PhysRevAccelBeams.25.080702} {\bibfield  {journal} {\bibinfo
  {journal} {Phys. Rev. Accel. Beams}\ }\textbf {\bibinfo {volume} {25}},\
  \bibinfo {pages} {080702} (\bibinfo {year} {2022})}\BibitemShut {NoStop}%
\bibitem [{\citenamefont {Hofmann}\ \emph {et~al.}(2018)\citenamefont
  {Hofmann}, \citenamefont {Boorman}, \citenamefont {Bosco}, \citenamefont
  {Gibson},\ and\ \citenamefont {Roncarolo}}]{Hofmann2018}%
  \BibitemOpen
  \bibfield  {author} {\bibinfo {author} {\bibfnamefont {T.}~\bibnamefont
  {Hofmann}}, \bibinfo {author} {\bibfnamefont {G.}~\bibnamefont {Boorman}},
  \bibinfo {author} {\bibfnamefont {A.}~\bibnamefont {Bosco}}, \bibinfo
  {author} {\bibfnamefont {S.}~\bibnamefont {Gibson}}, \ and\ \bibinfo {author}
  {\bibfnamefont {F.}~\bibnamefont {Roncarolo}},\ }\bibfield  {title} {\enquote
  {\bibinfo {title} {A low-power laserwire profile monitor for h- beams: Design
  and experimental results},}\ }\href {\doibase
  https://doi.org/10.1016/j.nima.2018.06.035} {\bibfield  {journal} {\bibinfo
  {journal} {Nuclear Instruments and Methods in Physics Research Section A:
  Accelerators, Spectrometers, Detectors and Associated Equipment}\ }\textbf
  {\bibinfo {volume} {903}},\ \bibinfo {pages} {140--146} (\bibinfo {year}
  {2018})}\BibitemShut {NoStop}%
\bibitem [{\citenamefont {Corner}\ \emph {et~al.}(2014)\citenamefont {Corner},
  \citenamefont {Aryshev}, \citenamefont {Blair}, \citenamefont {Boogert},
  \citenamefont {Karataev}, \citenamefont {Kruchinin}, \citenamefont {Nevay},
  \citenamefont {Terunuma}, \citenamefont {Urakawa},\ and\ \citenamefont
  {Walczak}}]{Corner2014}%
  \BibitemOpen
  \bibfield  {author} {\bibinfo {author} {\bibfnamefont {L.}~\bibnamefont
  {Corner}}, \bibinfo {author} {\bibfnamefont {A.}~\bibnamefont {Aryshev}},
  \bibinfo {author} {\bibfnamefont {G.}~\bibnamefont {Blair}}, \bibinfo
  {author} {\bibfnamefont {S.}~\bibnamefont {Boogert}}, \bibinfo {author}
  {\bibfnamefont {P.}~\bibnamefont {Karataev}}, \bibinfo {author}
  {\bibfnamefont {K.}~\bibnamefont {Kruchinin}}, \bibinfo {author}
  {\bibfnamefont {L.}~\bibnamefont {Nevay}}, \bibinfo {author} {\bibfnamefont
  {N.}~\bibnamefont {Terunuma}}, \bibinfo {author} {\bibfnamefont
  {J.}~\bibnamefont {Urakawa}}, \ and\ \bibinfo {author} {\bibfnamefont
  {R.}~\bibnamefont {Walczak}},\ }\bibfield  {title} {\enquote {\bibinfo
  {title} {Laserwire: A high resolution non-invasive beam profiling
  diagnostic},}\ }\href {\doibase https://doi.org/10.1016/j.nima.2013.10.043}
  {\bibfield  {journal} {\bibinfo  {journal} {Nuclear Instruments and Methods
  in Physics Research Section A: Accelerators, Spectrometers, Detectors and
  Associated Equipment}\ }\textbf {\bibinfo {volume} {740}},\ \bibinfo {pages}
  {226--228} (\bibinfo {year} {2014})},\ \bibinfo {note} {proceedings of the
  first European Advanced Accelerator Concepts Workshop 2013}\BibitemShut
  {NoStop}%
\bibitem [{\citenamefont {Anne}\ \emph {et~al.}(1993)\citenamefont {Anne},
  \citenamefont {Georget}, \citenamefont {Hue}, \citenamefont {Tribouillard},\
  and\ \citenamefont {{Luc Vignet}}}]{Anne1993}%
  \BibitemOpen
  \bibfield  {author} {\bibinfo {author} {\bibfnamefont {R.}~\bibnamefont
  {Anne}}, \bibinfo {author} {\bibfnamefont {Y.}~\bibnamefont {Georget}},
  \bibinfo {author} {\bibfnamefont {R.}~\bibnamefont {Hue}}, \bibinfo {author}
  {\bibfnamefont {C.}~\bibnamefont {Tribouillard}}, \ and\ \bibinfo {author}
  {\bibfnamefont {J.}~\bibnamefont {{Luc Vignet}}},\ }\bibfield  {title}
  {\enquote {\bibinfo {title} {A noninterceptive heavy ion beam profile monitor
  based on residual gas ionization},}\ }\href {\doibase
  https://doi.org/10.1016/0168-9002(93)90918-8} {\bibfield  {journal} {\bibinfo
   {journal} {Nuclear Instruments and Methods in Physics Research Section A:
  Accelerators, Spectrometers, Detectors and Associated Equipment}\ }\textbf
  {\bibinfo {volume} {329}},\ \bibinfo {pages} {21--28} (\bibinfo {year}
  {1993})}\BibitemShut {NoStop}%
\bibitem [{\citenamefont {Sandberg}\ \emph {et~al.}(2021)\citenamefont
  {Sandberg}, \citenamefont {Bertsche}, \citenamefont {Bodart}, \citenamefont
  {Gibson}, \citenamefont {Jensen}, \citenamefont {Levasseur}, \citenamefont
  {Satou}, \citenamefont {Schneider}, \citenamefont {Storey},\ and\
  \citenamefont {Veness}}]{Sandberg2021}%
  \BibitemOpen
  \bibfield  {author} {\bibinfo {author} {\bibfnamefont {H.}~\bibnamefont
  {Sandberg}}, \bibinfo {author} {\bibfnamefont {W.}~\bibnamefont {Bertsche}},
  \bibinfo {author} {\bibfnamefont {D.}~\bibnamefont {Bodart}}, \bibinfo
  {author} {\bibfnamefont {S.}~\bibnamefont {Gibson}}, \bibinfo {author}
  {\bibfnamefont {S.}~\bibnamefont {Jensen}}, \bibinfo {author} {\bibfnamefont
  {S.}~\bibnamefont {Levasseur}}, \bibinfo {author} {\bibfnamefont
  {K.}~\bibnamefont {Satou}}, \bibinfo {author} {\bibfnamefont
  {G.}~\bibnamefont {Schneider}}, \bibinfo {author} {\bibfnamefont
  {J.}~\bibnamefont {Storey}}, \ and\ \bibinfo {author} {\bibfnamefont
  {R.}~\bibnamefont {Veness}},\ }\bibfield  {title} {\enquote {\bibinfo {title}
  {Commissioning of timepix3 based beam gas ionisation profile monitors for the
  {CERN} proton synchrotron.}}\ }\href {\doibase
  10.18429/JACoW-IBIC2021-TUOA05} {\bibfield  {journal} {\bibinfo  {journal}
  {JACoW IBIC}\ }\textbf {\bibinfo {volume} {2021}},\ \bibinfo {pages}
  {172--175} (\bibinfo {year} {2021})}\BibitemShut {NoStop}%
\bibitem [{\citenamefont {Variola}, \citenamefont {Jung},\ and\ \citenamefont
  {Ferioli}(2007)}]{Variola2007}%
  \BibitemOpen
  \bibfield  {author} {\bibinfo {author} {\bibfnamefont {A.}~\bibnamefont
  {Variola}}, \bibinfo {author} {\bibfnamefont {R.}~\bibnamefont {Jung}}, \
  and\ \bibinfo {author} {\bibfnamefont {G.}~\bibnamefont {Ferioli}},\
  }\bibfield  {title} {\enquote {\bibinfo {title} {Characterization of a
  nondestructive beam profile monitor using luminescent emission},}\ }\href
  {\doibase 10.1103/PhysRevSTAB.10.122801} {\bibfield  {journal} {\bibinfo
  {journal} {Phys. Rev. ST Accel. Beams}\ }\textbf {\bibinfo {volume} {10}},\
  \bibinfo {pages} {122801} (\bibinfo {year} {2007})}\BibitemShut {NoStop}%
\bibitem [{\citenamefont {Salehilashkajani}\ \emph {et~al.}(2022)\citenamefont
  {Salehilashkajani}, \citenamefont {Zhang}, \citenamefont {Ady}, \citenamefont
  {Chritin}, \citenamefont {Forck}, \citenamefont {Glutting}, \citenamefont
  {Jones}, \citenamefont {Kersevan}, \citenamefont {Kumar}, \citenamefont
  {Lefevre}, \citenamefont {Marriott-Dodington}, \citenamefont {Mazzoni},
  \citenamefont {Papazoglou}, \citenamefont {Rossi}, \citenamefont {Schneider},
  \citenamefont {Sedlacek}, \citenamefont {Udrea}, \citenamefont {Veness},\
  and\ \citenamefont {Welsch}}]{Salehilashkajani2022}%
  \BibitemOpen
  \bibfield  {author} {\bibinfo {author} {\bibfnamefont {A.}~\bibnamefont
  {Salehilashkajani}}, \bibinfo {author} {\bibfnamefont {H.~D.}\ \bibnamefont
  {Zhang}}, \bibinfo {author} {\bibfnamefont {M.}~\bibnamefont {Ady}}, \bibinfo
  {author} {\bibfnamefont {N.}~\bibnamefont {Chritin}}, \bibinfo {author}
  {\bibfnamefont {P.}~\bibnamefont {Forck}}, \bibinfo {author} {\bibfnamefont
  {J.}~\bibnamefont {Glutting}}, \bibinfo {author} {\bibfnamefont {O.~R.}\
  \bibnamefont {Jones}}, \bibinfo {author} {\bibfnamefont {R.}~\bibnamefont
  {Kersevan}}, \bibinfo {author} {\bibfnamefont {N.}~\bibnamefont {Kumar}},
  \bibinfo {author} {\bibfnamefont {T.}~\bibnamefont {Lefevre}}, \bibinfo
  {author} {\bibfnamefont {T.}~\bibnamefont {Marriott-Dodington}}, \bibinfo
  {author} {\bibfnamefont {S.}~\bibnamefont {Mazzoni}}, \bibinfo {author}
  {\bibfnamefont {I.}~\bibnamefont {Papazoglou}}, \bibinfo {author}
  {\bibfnamefont {A.}~\bibnamefont {Rossi}}, \bibinfo {author} {\bibfnamefont
  {G.}~\bibnamefont {Schneider}}, \bibinfo {author} {\bibfnamefont
  {O.}~\bibnamefont {Sedlacek}}, \bibinfo {author} {\bibfnamefont
  {S.}~\bibnamefont {Udrea}}, \bibinfo {author} {\bibfnamefont
  {R.}~\bibnamefont {Veness}}, \ and\ \bibinfo {author} {\bibfnamefont {C.~P.}\
  \bibnamefont {Welsch}},\ }\bibfield  {title} {\enquote {\bibinfo {title} {A
  gas curtain beam profile monitor using beam induced fluorescence for high
  intensity charged particle beams},}\ }\href {\doibase 10.1063/5.0085491}
  {\bibfield  {journal} {\bibinfo  {journal} {Applied Physics Letters}\
  }\textbf {\bibinfo {volume} {120}},\ \bibinfo {pages} {174101} (\bibinfo
  {year} {2022})}\BibitemShut {NoStop}%
\bibitem [{\citenamefont {Maesaka}\ \emph {et~al.}(2012)\citenamefont
  {Maesaka}, \citenamefont {Ego}, \citenamefont {Inoue}, \citenamefont
  {Matsubara}, \citenamefont {Ohshima}, \citenamefont {Shintake},\ and\
  \citenamefont {Otake}}]{Maesaka2012}%
  \BibitemOpen
  \bibfield  {author} {\bibinfo {author} {\bibfnamefont {H.}~\bibnamefont
  {Maesaka}}, \bibinfo {author} {\bibfnamefont {H.}~\bibnamefont {Ego}},
  \bibinfo {author} {\bibfnamefont {S.}~\bibnamefont {Inoue}}, \bibinfo
  {author} {\bibfnamefont {S.}~\bibnamefont {Matsubara}}, \bibinfo {author}
  {\bibfnamefont {T.}~\bibnamefont {Ohshima}}, \bibinfo {author} {\bibfnamefont
  {T.}~\bibnamefont {Shintake}}, \ and\ \bibinfo {author} {\bibfnamefont
  {Y.}~\bibnamefont {Otake}},\ }\bibfield  {title} {\enquote {\bibinfo {title}
  {Sub-micron resolution rf cavity beam position monitor system at the sacla
  xfel facility},}\ }\href {\doibase
  https://doi.org/10.1016/j.nima.2012.08.088} {\bibfield  {journal} {\bibinfo
  {journal} {Nuclear Instruments and Methods in Physics Research Section A:
  Accelerators, Spectrometers, Detectors and Associated Equipment}\ }\textbf
  {\bibinfo {volume} {696}},\ \bibinfo {pages} {66--74} (\bibinfo {year}
  {2012})}\BibitemShut {NoStop}%
\bibitem [{\citenamefont {Budker}\ and\ \citenamefont
  {Jackson~Kimball}(2013)}]{Budker2013}%
  \BibitemOpen
  \bibinfo {editor} {\bibfnamefont {D.}~\bibnamefont {Budker}}\ and\ \bibinfo
  {editor} {\bibfnamefont {D.~F.}\ \bibnamefont {Jackson~Kimball}},\ eds.,\
  \href {\doibase 10.1017/CBO9780511846380} {\emph {\bibinfo {title} {{Optical
  Magnetometry}}}}\ (\bibinfo  {publisher} {Cambridge University Press},\
  \bibinfo {address} {Cambridge},\ \bibinfo {year} {2013})\BibitemShut
  {NoStop}%
\bibitem [{\citenamefont {Weis}, \citenamefont {Bison},\ and\ \citenamefont
  {Gruji{\'{c}}}(2017)}]{Weis2017}%
  \BibitemOpen
  \bibfield  {author} {\bibinfo {author} {\bibfnamefont {A.}~\bibnamefont
  {Weis}}, \bibinfo {author} {\bibfnamefont {G.}~\bibnamefont {Bison}}, \ and\
  \bibinfo {author} {\bibfnamefont {Z.~D.}\ \bibnamefont {Gruji{\'{c}}}},\
  }\bibfield  {title} {\enquote {\bibinfo {title} {{Magnetic resonance based
  atomic magnetometers}},}\ }in\ \href {\doibase 10.1007/978-3-319-34070-8}
  {\emph {\bibinfo {booktitle} {High Sensitivity Magnetometers. Smart Sensors,
  Measurement and Instrumentation}}},\ Vol.~\bibinfo {volume} {19},\ \bibinfo
  {editor} {edited by\ \bibinfo {editor} {\bibfnamefont {A.}~\bibnamefont
  {Grosz}}, \bibinfo {editor} {\bibfnamefont {M.}~\bibnamefont {Haji-Sheikh}},
  \ and\ \bibinfo {editor} {\bibfnamefont {S.}~\bibnamefont {Mukhopadhyay}}}\
  (\bibinfo  {publisher} {Springer},\ \bibinfo {address} {Cham},\ \bibinfo
  {year} {2017})\ pp.\ \bibinfo {pages} {361--424}\BibitemShut {NoStop}%
\bibitem [{\citenamefont {Fu}\ \emph {et~al.}(2020)\citenamefont {Fu},
  \citenamefont {Iwata}, \citenamefont {Wickenbrock},\ and\ \citenamefont
  {Budker}}]{Fu2020}%
  \BibitemOpen
  \bibfield  {author} {\bibinfo {author} {\bibfnamefont {K.~M.~C.}\
  \bibnamefont {Fu}}, \bibinfo {author} {\bibfnamefont {G.~Z.}\ \bibnamefont
  {Iwata}}, \bibinfo {author} {\bibfnamefont {A.}~\bibnamefont {Wickenbrock}},
  \ and\ \bibinfo {author} {\bibfnamefont {D.}~\bibnamefont {Budker}},\
  }\bibfield  {title} {\enquote {\bibinfo {title} {{Sensitive magnetometry in
  challenging environments}},}\ }\href {\doibase 10.1116/5.0025186} {\bibfield
  {journal} {\bibinfo  {journal} {AVS Quantum Science}\ }\textbf {\bibinfo
  {volume} {2}},\ \bibinfo {pages} {044702} (\bibinfo {year}
  {2020})}\BibitemShut {NoStop}%
\bibitem [{Note1()}]{Note1}%
  \BibitemOpen
  \bibinfo {note} {We calculate a Bethe energy loss of 0.1 eV/m at a Rb density
  of $10^{12}~\protect \mathrm {atoms/cm}^3$ for 20 keV electrons. At 1 GeV,
  the energy loss is 0.04 eV/m.}\BibitemShut {Stop}%
\bibitem [{\citenamefont {Balabas}\ \emph {et~al.}(2010)\citenamefont
  {Balabas}, \citenamefont {Karaulanov}, \citenamefont {Ledbetter},\ and\
  \citenamefont {Budker}}]{BalabasPRL2010}%
  \BibitemOpen
  \bibfield  {author} {\bibinfo {author} {\bibfnamefont {M.~V.}\ \bibnamefont
  {Balabas}}, \bibinfo {author} {\bibfnamefont {T.}~\bibnamefont {Karaulanov}},
  \bibinfo {author} {\bibfnamefont {M.~P.}\ \bibnamefont {Ledbetter}}, \ and\
  \bibinfo {author} {\bibfnamefont {D.}~\bibnamefont {Budker}},\ }\bibfield
  {title} {\enquote {\bibinfo {title} {Polarized alkali-metal vapor with
  minute-long transverse spin-relaxation time},}\ }\href {\doibase
  10.1103/PhysRevLett.105.070801} {\bibfield  {journal} {\bibinfo  {journal}
  {Phys. Rev. Lett.}\ }\textbf {\bibinfo {volume} {105}},\ \bibinfo {pages}
  {070801} (\bibinfo {year} {2010})}\BibitemShut {NoStop}%
\bibitem [{\citenamefont {Rosner}\ \emph {et~al.}(2022)\citenamefont {Rosner},
  \citenamefont {Beck}, \citenamefont {Fierlinger}, \citenamefont {Filter},
  \citenamefont {Klau}, \citenamefont {Kuchler}, \citenamefont
  {R{\"{o}}{\ss}ner}, \citenamefont {Sturm}, \citenamefont {Wurm},\ and\
  \citenamefont {Sun}}]{Rosner2022}%
  \BibitemOpen
  \bibfield  {author} {\bibinfo {author} {\bibfnamefont {M.}~\bibnamefont
  {Rosner}}, \bibinfo {author} {\bibfnamefont {D.}~\bibnamefont {Beck}},
  \bibinfo {author} {\bibfnamefont {P.}~\bibnamefont {Fierlinger}}, \bibinfo
  {author} {\bibfnamefont {H.}~\bibnamefont {Filter}}, \bibinfo {author}
  {\bibfnamefont {C.}~\bibnamefont {Klau}}, \bibinfo {author} {\bibfnamefont
  {F.}~\bibnamefont {Kuchler}}, \bibinfo {author} {\bibfnamefont
  {P.}~\bibnamefont {R{\"{o}}{\ss}ner}}, \bibinfo {author} {\bibfnamefont
  {M.}~\bibnamefont {Sturm}}, \bibinfo {author} {\bibfnamefont
  {D.}~\bibnamefont {Wurm}}, \ and\ \bibinfo {author} {\bibfnamefont
  {Z.}~\bibnamefont {Sun}},\ }\bibfield  {title} {\enquote {\bibinfo {title}
  {{A highly drift-stable atomic magnetometer for fundamental physics
  experiments}},}\ }\href {\doibase 10.1063/5.0083854} {\bibfield  {journal}
  {\bibinfo  {journal} {Applied Physics Letters}\ }\textbf {\bibinfo {volume}
  {120}},\ \bibinfo {pages} {161102} (\bibinfo {year} {2022})}\BibitemShut
  {NoStop}%
\bibitem [{\citenamefont {Lucivero}\ \emph {et~al.}(2021)\citenamefont
  {Lucivero}, \citenamefont {Lee}, \citenamefont {Dural},\ and\ \citenamefont
  {Romalis}}]{Lucivero2021}%
  \BibitemOpen
  \bibfield  {author} {\bibinfo {author} {\bibfnamefont {V.~G.}\ \bibnamefont
  {Lucivero}}, \bibinfo {author} {\bibfnamefont {W.}~\bibnamefont {Lee}},
  \bibinfo {author} {\bibfnamefont {N.}~\bibnamefont {Dural}}, \ and\ \bibinfo
  {author} {\bibfnamefont {M.~V.}\ \bibnamefont {Romalis}},\ }\bibfield
  {title} {\enquote {\bibinfo {title} {{Femtotesla direct magnetic gradiometer
  using a single multipass cell}},}\ }\href {\doibase
  10.1103/PhysRevApplied.15.014004} {\bibfield  {journal} {\bibinfo  {journal}
  {Physical Review Applied}\ }\textbf {\bibinfo {volume} {15}},\ \bibinfo
  {pages} {014004} (\bibinfo {year} {2021})}\BibitemShut {NoStop}%
\bibitem [{\citenamefont {Budker}\ \emph {et~al.}(2002)\citenamefont {Budker},
  \citenamefont {Gawlik}, \citenamefont {Kimball}, \citenamefont {Rochester},
  \citenamefont {Yashchuk},\ and\ \citenamefont {Weis}}]{Budker_RMP2002}%
  \BibitemOpen
  \bibfield  {author} {\bibinfo {author} {\bibfnamefont {D.}~\bibnamefont
  {Budker}}, \bibinfo {author} {\bibfnamefont {W.}~\bibnamefont {Gawlik}},
  \bibinfo {author} {\bibfnamefont {D.~F.}\ \bibnamefont {Kimball}}, \bibinfo
  {author} {\bibfnamefont {S.~M.}\ \bibnamefont {Rochester}}, \bibinfo {author}
  {\bibfnamefont {V.~V.}\ \bibnamefont {Yashchuk}}, \ and\ \bibinfo {author}
  {\bibfnamefont {A.}~\bibnamefont {Weis}},\ }\bibfield  {title} {\enquote
  {\bibinfo {title} {Resonant nonlinear magneto-optical effects in atoms},}\
  }\href {\doibase 10.1103/RevModPhys.74.1153} {\bibfield  {journal} {\bibinfo
  {journal} {Rev. Mod. Phys.}\ }\textbf {\bibinfo {volume} {74}},\ \bibinfo
  {pages} {1153} (\bibinfo {year} {2002})}\BibitemShut {NoStop}%
\bibitem [{\citenamefont {Anisimov}\ \emph {et~al.}(2003)\citenamefont
  {Anisimov}, \citenamefont {Akhmedzhanov}, \citenamefont {Zelensky},\ and\
  \citenamefont {Kuznetsova}}]{Anisimov2003}%
  \BibitemOpen
  \bibfield  {author} {\bibinfo {author} {\bibfnamefont {P.~M.}\ \bibnamefont
  {Anisimov}}, \bibinfo {author} {\bibfnamefont {R.~A.}\ \bibnamefont
  {Akhmedzhanov}}, \bibinfo {author} {\bibfnamefont {I.~V.}\ \bibnamefont
  {Zelensky}}, \ and\ \bibinfo {author} {\bibfnamefont {E.~A.}\ \bibnamefont
  {Kuznetsova}},\ }\bibfield  {title} {\enquote {\bibinfo {title} {Influence of
  transverse magnetic fields and depletion of working levels on the nonlinear
  resonance faraday effect},}\ }\href {\doibase 10.1134/1.1633944} {\bibfield
  {journal} {\bibinfo  {journal} {Journal of Experimental and Theoretical
  Physics}\ }\textbf {\bibinfo {volume} {97}},\ \bibinfo {pages} {868--874}
  (\bibinfo {year} {2003})}\BibitemShut {NoStop}%
\bibitem [{\citenamefont {Zhou}\ \emph {et~al.}(2018)\citenamefont {Zhou},
  \citenamefont {Zhu}, \citenamefont {Li}, \citenamefont {Hagley},\ and\
  \citenamefont {Deng}}]{Zhuarxiv1804.08194}%
  \BibitemOpen
  \bibfield  {author} {\bibinfo {author} {\bibfnamefont {F.}~\bibnamefont
  {Zhou}}, \bibinfo {author} {\bibfnamefont {E.~Y.}\ \bibnamefont {Zhu}},
  \bibinfo {author} {\bibfnamefont {Y.~L.}\ \bibnamefont {Li}}, \bibinfo
  {author} {\bibfnamefont {E.~W.}\ \bibnamefont {Hagley}}, \ and\ \bibinfo
  {author} {\bibfnamefont {L.}~\bibnamefont {Deng}},\ }\href {\doibase
  10.48550/ARXIV.1804.08194} {\enquote {\bibinfo {title} {Nanotesla-level,
  shield-less, field-compensation-free, wave-mixing-enhanced body-temperature
  atomic magnetometry for biomagnetism},}\ } (\bibinfo {year}
  {2018})\BibitemShut {NoStop}%
\bibitem [{\citenamefont {Zhu}\ \emph {et~al.}(2021)\citenamefont {Zhu},
  \citenamefont {Guan}, \citenamefont {Zhou}, \citenamefont {Zhu},\ and\
  \citenamefont {Li}}]{Zhu:21}%
  \BibitemOpen
  \bibfield  {author} {\bibinfo {author} {\bibfnamefont {C.~J.}\ \bibnamefont
  {Zhu}}, \bibinfo {author} {\bibfnamefont {J.}~\bibnamefont {Guan}}, \bibinfo
  {author} {\bibfnamefont {F.}~\bibnamefont {Zhou}}, \bibinfo {author}
  {\bibfnamefont {E.~Y.}\ \bibnamefont {Zhu}}, \ and\ \bibinfo {author}
  {\bibfnamefont {Y.}~\bibnamefont {Li}},\ }\bibfield  {title} {\enquote
  {\bibinfo {title} {Giant magneto-optical rotation effect in rubidium vapor
  measured with a low-cost detection system},}\ }\href {\doibase
  10.1364/OSAC.435754} {\bibfield  {journal} {\bibinfo  {journal} {OSA
  Continuum}\ }\textbf {\bibinfo {volume} {4}},\ \bibinfo {pages} {2527--2534}
  (\bibinfo {year} {2021})}\BibitemShut {NoStop}%
\bibitem [{\citenamefont {Wolfgramm}\ \emph {et~al.}(2010)\citenamefont
  {Wolfgramm}, \citenamefont {Cer{\`{e}}}, \citenamefont {Beduini},
  \citenamefont {Predojevi{\'{c}}}, \citenamefont {Koschorreck},\ and\
  \citenamefont {Mitchell}}]{Wolfgramm2010}%
  \BibitemOpen
  \bibfield  {author} {\bibinfo {author} {\bibfnamefont {F.}~\bibnamefont
  {Wolfgramm}}, \bibinfo {author} {\bibfnamefont {A.}~\bibnamefont
  {Cer{\`{e}}}}, \bibinfo {author} {\bibfnamefont {F.~A.}\ \bibnamefont
  {Beduini}}, \bibinfo {author} {\bibfnamefont {A.}~\bibnamefont
  {Predojevi{\'{c}}}}, \bibinfo {author} {\bibfnamefont {M.}~\bibnamefont
  {Koschorreck}}, \ and\ \bibinfo {author} {\bibfnamefont {M.~W.}\ \bibnamefont
  {Mitchell}},\ }\bibfield  {title} {\enquote {\bibinfo {title}
  {{Squeezed-light optical magnetometry}},}\ }\href {\doibase
  10.1103/PhysRevLett.105.053601} {\bibfield  {journal} {\bibinfo  {journal}
  {Physical Review Letters}\ }\textbf {\bibinfo {volume} {105}},\ \bibinfo
  {pages} {053601} (\bibinfo {year} {2010})}\BibitemShut {NoStop}%
\bibitem [{\citenamefont {Horrom}\ \emph {et~al.}(2012)\citenamefont {Horrom},
  \citenamefont {Singh}, \citenamefont {Dowling},\ and\ \citenamefont
  {Mikhailov}}]{Horrom2012}%
  \BibitemOpen
  \bibfield  {author} {\bibinfo {author} {\bibfnamefont {T.}~\bibnamefont
  {Horrom}}, \bibinfo {author} {\bibfnamefont {R.}~\bibnamefont {Singh}},
  \bibinfo {author} {\bibfnamefont {J.~P.}\ \bibnamefont {Dowling}}, \ and\
  \bibinfo {author} {\bibfnamefont {E.~E.}\ \bibnamefont {Mikhailov}},\
  }\bibfield  {title} {\enquote {\bibinfo {title} {{Quantum-enhanced
  magnetometer with low-frequency squeezing}},}\ }\href {\doibase
  10.1103/PHYSREVA.86.023803/FIGURES/8/MEDIUM} {\bibfield  {journal} {\bibinfo
  {journal} {Physical Review A - Atomic, Molecular, and Optical Physics}\
  }\textbf {\bibinfo {volume} {86}},\ \bibinfo {pages} {023803} (\bibinfo
  {year} {2012})}\BibitemShut {NoStop}%
\bibitem [{\citenamefont {Otterstrom}, \citenamefont {Pooser},\ and\
  \citenamefont {Lawrie}(2014)}]{OtterstromOL2014}%
  \BibitemOpen
  \bibfield  {author} {\bibinfo {author} {\bibfnamefont {N.}~\bibnamefont
  {Otterstrom}}, \bibinfo {author} {\bibfnamefont {R.~C.}\ \bibnamefont
  {Pooser}}, \ and\ \bibinfo {author} {\bibfnamefont {B.~J.}\ \bibnamefont
  {Lawrie}},\ }\bibfield  {title} {\enquote {\bibinfo {title} {Nonlinear
  optical magnetometry with accessible in situ optical squeezing},}\ }\href
  {\doibase 10.1364/OL.39.006533} {\bibfield  {journal} {\bibinfo  {journal}
  {Opt. Lett.}\ }\textbf {\bibinfo {volume} {39}},\ \bibinfo {pages}
  {6533--6536} (\bibinfo {year} {2014})}\BibitemShut {NoStop}%
\bibitem [{\citenamefont {Bai}\ \emph {et~al.}(2021)\citenamefont {Bai},
  \citenamefont {Wen}, \citenamefont {Yang}, \citenamefont {Zhang},
  \citenamefont {He}, \citenamefont {Wang},\ and\ \citenamefont
  {Wang}}]{BaiJOpt2021}%
  \BibitemOpen
  \bibfield  {author} {\bibinfo {author} {\bibfnamefont {L.}~\bibnamefont
  {Bai}}, \bibinfo {author} {\bibfnamefont {X.}~\bibnamefont {Wen}}, \bibinfo
  {author} {\bibfnamefont {Y.}~\bibnamefont {Yang}}, \bibinfo {author}
  {\bibfnamefont {L.}~\bibnamefont {Zhang}}, \bibinfo {author} {\bibfnamefont
  {J.}~\bibnamefont {He}}, \bibinfo {author} {\bibfnamefont {Y.}~\bibnamefont
  {Wang}}, \ and\ \bibinfo {author} {\bibfnamefont {J.}~\bibnamefont {Wang}},\
  }\bibfield  {title} {\enquote {\bibinfo {title} {Quantum-enhanced rubidium
  atomic magnetometer based on faraday rotation via 795 nm stokes operator
  squeezed light},}\ }\href {\doibase 10.1088/2040-8986/ac1b7c} {\bibfield
  {journal} {\bibinfo  {journal} {Journal of Optics}\ }\textbf {\bibinfo
  {volume} {23}},\ \bibinfo {pages} {085202} (\bibinfo {year}
  {2021})}\BibitemShut {NoStop}%
\bibitem [{\citenamefont {Li}\ and\ \citenamefont {Novikova}(2022)}]{Li2022}%
  \BibitemOpen
  \bibfield  {author} {\bibinfo {author} {\bibfnamefont {J.}~\bibnamefont
  {Li}}\ and\ \bibinfo {author} {\bibfnamefont {I.}~\bibnamefont {Novikova}},\
  }\bibfield  {title} {\enquote {\bibinfo {title} {Improving sensitivity of an
  amplitude-modulated magneto-optical atomic magnetometer using squeezed
  light},}\ }\href {\doibase 10.1364/JOSAB.471677} {\bibfield  {journal}
  {\bibinfo  {journal} {J. Opt. Soc. Am. B}\ }\textbf {\bibinfo {volume}
  {39}},\ \bibinfo {pages} {2998--3003} (\bibinfo {year} {2022})}\BibitemShut
  {NoStop}%
\bibitem [{\citenamefont {Wu}\ \emph {et~al.}(2023)\citenamefont {Wu},
  \citenamefont {Bao}, \citenamefont {Guo}, \citenamefont {Chen}, \citenamefont
  {Du}, \citenamefont {Shi}, \citenamefont {Yang}, \citenamefont {Chen},\ and\
  \citenamefont {Zhang}}]{WuSA2023}%
  \BibitemOpen
  \bibfield  {author} {\bibinfo {author} {\bibfnamefont {S.}~\bibnamefont
  {Wu}}, \bibinfo {author} {\bibfnamefont {G.}~\bibnamefont {Bao}}, \bibinfo
  {author} {\bibfnamefont {J.}~\bibnamefont {Guo}}, \bibinfo {author}
  {\bibfnamefont {J.}~\bibnamefont {Chen}}, \bibinfo {author} {\bibfnamefont
  {W.}~\bibnamefont {Du}}, \bibinfo {author} {\bibfnamefont {M.}~\bibnamefont
  {Shi}}, \bibinfo {author} {\bibfnamefont {P.}~\bibnamefont {Yang}}, \bibinfo
  {author} {\bibfnamefont {L.}~\bibnamefont {Chen}}, \ and\ \bibinfo {author}
  {\bibfnamefont {W.}~\bibnamefont {Zhang}},\ }\bibfield  {title} {\enquote
  {\bibinfo {title} {Quantum magnetic gradiometer with entangled twin light
  beams},}\ }\href {\doibase 10.1126/sciadv.adg1760} {\bibfield  {journal}
  {\bibinfo  {journal} {Science Advances}\ }\textbf {\bibinfo {volume} {9}},\
  \bibinfo {pages} {eadg1760} (\bibinfo {year} {2023})}\BibitemShut {NoStop}%
\bibitem [{\citenamefont {Boyer}\ \emph {et~al.}(2007)\citenamefont {Boyer},
  \citenamefont {McCormick}, \citenamefont {Arimondo},\ and\ \citenamefont
  {Lett}}]{BoyerLett_PRL2007}%
  \BibitemOpen
  \bibfield  {author} {\bibinfo {author} {\bibfnamefont {V.}~\bibnamefont
  {Boyer}}, \bibinfo {author} {\bibfnamefont {C.~F.}\ \bibnamefont
  {McCormick}}, \bibinfo {author} {\bibfnamefont {E.}~\bibnamefont {Arimondo}},
  \ and\ \bibinfo {author} {\bibfnamefont {P.~D.}\ \bibnamefont {Lett}},\
  }\bibfield  {title} {\enquote {\bibinfo {title} {Ultraslow propagation of
  matched pulses by four-wave mixing in an atomic vapor},}\ }\href {\doibase
  10.1103/PhysRevLett.99.143601} {\bibfield  {journal} {\bibinfo  {journal}
  {Phys. Rev. Lett.}\ }\textbf {\bibinfo {volume} {99}},\ \bibinfo {pages}
  {143601} (\bibinfo {year} {2007})}\BibitemShut {NoStop}%
\bibitem [{\citenamefont {Thaicharoen}\ \emph {et~al.}(2019)\citenamefont
  {Thaicharoen}, \citenamefont {Moore}, \citenamefont {Anderson}, \citenamefont
  {Powel}, \citenamefont {Peterson},\ and\ \citenamefont
  {Raithel}}]{Raithel_PRA2019}%
  \BibitemOpen
  \bibfield  {author} {\bibinfo {author} {\bibfnamefont {N.}~\bibnamefont
  {Thaicharoen}}, \bibinfo {author} {\bibfnamefont {K.~R.}\ \bibnamefont
  {Moore}}, \bibinfo {author} {\bibfnamefont {D.~A.}\ \bibnamefont {Anderson}},
  \bibinfo {author} {\bibfnamefont {R.~C.}\ \bibnamefont {Powel}}, \bibinfo
  {author} {\bibfnamefont {E.}~\bibnamefont {Peterson}}, \ and\ \bibinfo
  {author} {\bibfnamefont {G.}~\bibnamefont {Raithel}},\ }\bibfield  {title}
  {\enquote {\bibinfo {title} {Electromagnetically induced transparency,
  absorption, and microwave-field sensing in a rb vapor cell},}\ }\href
  {\doibase 10.1103/PhysRevA.100.063427} {\bibfield  {journal} {\bibinfo
  {journal} {Phys. Rev. A}\ }\textbf {\bibinfo {volume} {100}},\ \bibinfo
  {pages} {063427} (\bibinfo {year} {2019})}\BibitemShut {NoStop}%
\bibitem [{\citenamefont {Dietsche}\ \emph {et~al.}(2019)\citenamefont
  {Dietsche}, \citenamefont {Larrouy}, \citenamefont {Haroche}, \citenamefont
  {Raimond}, \citenamefont {Brune},\ and\ \citenamefont
  {Gleyzes}}]{Haroche_NatPhys2019}%
  \BibitemOpen
  \bibfield  {author} {\bibinfo {author} {\bibfnamefont {E.~K.}\ \bibnamefont
  {Dietsche}}, \bibinfo {author} {\bibfnamefont {A.}~\bibnamefont {Larrouy}},
  \bibinfo {author} {\bibfnamefont {S.}~\bibnamefont {Haroche}}, \bibinfo
  {author} {\bibfnamefont {J.~M.}\ \bibnamefont {Raimond}}, \bibinfo {author}
  {\bibfnamefont {M.}~\bibnamefont {Brune}}, \ and\ \bibinfo {author}
  {\bibfnamefont {S.}~\bibnamefont {Gleyzes}},\ }\bibfield  {title} {\enquote
  {\bibinfo {title} {High-sensitivity magnetometry with a single atom in a
  superposition of two circular rydberg states},}\ }\href {\doibase
  10.1038/s41567-018-0405-4} {\bibfield  {journal} {\bibinfo  {journal} {Nature
  Phys.}\ }\textbf {\bibinfo {volume} {15}},\ \bibinfo {pages} {326} (\bibinfo
  {year} {2019})}\BibitemShut {NoStop}%
\bibitem [{\citenamefont {Jerkins}\ \emph {et~al.}(2010)\citenamefont
  {Jerkins}, \citenamefont {Klein}, \citenamefont {Majors}, \citenamefont
  {Robicheaux},\ and\ \citenamefont {Raizen}}]{Raizen_NJP2010}%
  \BibitemOpen
  \bibfield  {author} {\bibinfo {author} {\bibfnamefont {M.}~\bibnamefont
  {Jerkins}}, \bibinfo {author} {\bibfnamefont {J.~P.}\ \bibnamefont {Klein}},
  \bibinfo {author} {\bibfnamefont {J.~H.}\ \bibnamefont {Majors}}, \bibinfo
  {author} {\bibfnamefont {F.}~\bibnamefont {Robicheaux}}, \ and\ \bibinfo
  {author} {\bibfnamefont {M.~G.}\ \bibnamefont {Raizen}},\ }\bibfield  {title}
  {\enquote {\bibinfo {title} {Using cold atoms to measure neutrino mass},}\
  }\href {\doibase 10.1088/1367-2630/12/4/043022} {\bibfield  {journal}
  {\bibinfo  {journal} {New J. Phys.}\ }\textbf {\bibinfo {volume} {12}},\
  \bibinfo {pages} {043022} (\bibinfo {year} {2010})}\BibitemShut {NoStop}%
\bibitem [{\citenamefont {Kawasaki}(2023)}]{Kawasaki_PhysRevRes2023}%
  \BibitemOpen
  \bibfield  {author} {\bibinfo {author} {\bibfnamefont {A.}~\bibnamefont
  {Kawasaki}},\ }\bibfield  {title} {\enquote {\bibinfo {title} {Tracking a
  nonrelativistic charge with an array of rydberg atoms},}\ }\href {\doibase
  10.1103/PhysRevResearch.5.043178} {\bibfield  {journal} {\bibinfo  {journal}
  {Phys. Rev. Res.}\ }\textbf {\bibinfo {volume} {5}},\ \bibinfo {pages}
  {043178} (\bibinfo {year} {2023})}\BibitemShut {NoStop}%
\bibitem [{\citenamefont {Ramaswamy}\ and\ \citenamefont
  {Malinovskaya}(2022)}]{Ramaswamy2022}%
  \BibitemOpen
  \bibfield  {author} {\bibinfo {author} {\bibfnamefont {A.}~\bibnamefont
  {Ramaswamy}}\ and\ \bibinfo {author} {\bibfnamefont {S.~A.}\ \bibnamefont
  {Malinovskaya}},\ }\enquote {\bibinfo {title} {Control with eit: High energy
  charged particle detection},}\ in\ \href {\doibase
  10.1007/978-3-030-93460-6_12} {\emph {\bibinfo {booktitle} {Progress in
  Nanoscale and Low-Dimensional Materials and Devices: Properties, Synthesis,
  Characterization, Modelling and Applications}}},\ \bibinfo {editor} {edited
  by\ \bibinfo {editor} {\bibfnamefont {H.}~\bibnamefont {{\"U}nl{\"u}}}\ and\
  \bibinfo {editor} {\bibfnamefont {N.~J.~M.}\ \bibnamefont {Horing}}}\
  (\bibinfo  {publisher} {Springer International Publishing},\ \bibinfo
  {address} {Cham},\ \bibinfo {year} {2022})\ pp.\ \bibinfo {pages}
  {363--392}\BibitemShut {NoStop}%
\end{thebibliography}%


\begin{thebibliography}{7}%
\makeatletter
\providecommand \@ifxundefined [1]{%
 \@ifx{#1\undefined}
}%
\providecommand \@ifnum [1]{%
 \ifnum #1\expandafter \@firstoftwo
 \else \expandafter \@secondoftwo
 \fi
}%
\providecommand \@ifx [1]{%
 \ifx #1\expandafter \@firstoftwo
 \else \expandafter \@secondoftwo
 \fi
}%
\providecommand \natexlab [1]{#1}%
\providecommand \enquote  [1]{``#1''}%
\providecommand \bibnamefont  [1]{#1}%
\providecommand \bibfnamefont [1]{#1}%
\providecommand \citenamefont [1]{#1}%
\providecommand \href@noop [0]{\@secondoftwo}%
\providecommand \href [0]{\begingroup \@sanitize@url \@href}%
\providecommand \@href[1]{\@@startlink{#1}\@@href}%
\providecommand \@@href[1]{\endgroup#1\@@endlink}%
\providecommand \@sanitize@url [0]{\catcode `\\12\catcode `\$12\catcode
  `\&12\catcode `\#12\catcode `\^12\catcode `\_12\catcode `\%12\relax}%
\providecommand \@@startlink[1]{}%
\providecommand \@@endlink[0]{}%
\providecommand \url  [0]{\begingroup\@sanitize@url \@url }%
\providecommand \@url [1]{\endgroup\@href {#1}{\urlprefix }}%
\providecommand \urlprefix  [0]{URL }%
\providecommand \Eprint [0]{\href }%
\providecommand \doibase [0]{http://dx.doi.org/}%
\providecommand \selectlanguage [0]{\@gobble}%
\providecommand \bibinfo  [0]{\@secondoftwo}%
\providecommand \bibfield  [0]{\@secondoftwo}%
\providecommand \translation [1]{[#1]}%
\providecommand \BibitemOpen [0]{}%
\providecommand \bibitemStop [0]{}%
\providecommand \bibitemNoStop [0]{.\EOS\space}%
\providecommand \EOS [0]{\spacefactor3000\relax}%
\providecommand \BibitemShut  [1]{\csname bibitem#1\endcsname}%
\let\auto@bib@innerbib\@empty
\bibitem [{\citenamefont {Budker}\ \emph {et~al.}(2002)\citenamefont {Budker},
  \citenamefont {Gawlik}, \citenamefont {Kimball}, \citenamefont {Rochester},
  \citenamefont {Yashchuk},\ and\ \citenamefont {Weis}}]{Budker_RMP2002}%
  \BibitemOpen
  \bibfield  {author} {\bibinfo {author} {\bibfnamefont {D.}~\bibnamefont
  {Budker}}, \bibinfo {author} {\bibfnamefont {W.}~\bibnamefont {Gawlik}},
  \bibinfo {author} {\bibfnamefont {D.~F.}\ \bibnamefont {Kimball}}, \bibinfo
  {author} {\bibfnamefont {S.~M.}\ \bibnamefont {Rochester}}, \bibinfo {author}
  {\bibfnamefont {V.~V.}\ \bibnamefont {Yashchuk}}, \ and\ \bibinfo {author}
  {\bibfnamefont {A.}~\bibnamefont {Weis}},\ }\bibfield  {title} {\enquote
  {\bibinfo {title} {Resonant nonlinear magneto-optical effects in atoms},}\
  }\href {\doibase 10.1103/RevModPhys.74.1153} {\bibfield  {journal} {\bibinfo
  {journal} {Rev. Mod. Phys.}\ }\textbf {\bibinfo {volume} {74}},\ \bibinfo
  {pages} {1153} (\bibinfo {year} {2002})}\BibitemShut {NoStop}%
\bibitem [{\citenamefont {Matsko}\ \emph {et~al.}(2003)\citenamefont {Matsko},
  \citenamefont {Novikova}, \citenamefont {Zubairy},\ and\ \citenamefont
  {Welch}}]{MatskoPRA2003}%
  \BibitemOpen
  \bibfield  {author} {\bibinfo {author} {\bibfnamefont {A.~B.}\ \bibnamefont
  {Matsko}}, \bibinfo {author} {\bibfnamefont {I.}~\bibnamefont {Novikova}},
  \bibinfo {author} {\bibfnamefont {M.~S.}\ \bibnamefont {Zubairy}}, \ and\
  \bibinfo {author} {\bibfnamefont {G.~R.}\ \bibnamefont {Welch}},\ }\bibfield
  {title} {\enquote {\bibinfo {title} {Nonlinear magneto-optical rotation of
  elliptically polarized light},}\ }\href {\doibase 10.1103/PhysRevA.67.043805}
  {\bibfield  {journal} {\bibinfo  {journal} {Phys. Rev. A}\ }\textbf {\bibinfo
  {volume} {67}},\ \bibinfo {pages} {043805} (\bibinfo {year}
  {2003})}\BibitemShut {NoStop}%
\bibitem [{\citenamefont {Harris}(1997)}]{harris'97pt}%
  \BibitemOpen
  \bibfield  {author} {\bibinfo {author} {\bibfnamefont {S.~E.}\ \bibnamefont
  {Harris}},\ }\bibfield  {title} {\enquote {\bibinfo {title}
  {Electromagnetically induced transparency},}\ }\href {\doibase
  10.1063/1.881806} {\bibfield  {journal} {\bibinfo  {journal} {Physics Today}\
  }\textbf {\bibinfo {volume} {50}},\ \bibinfo {pages} {36} (\bibinfo {year}
  {1997})}\BibitemShut {NoStop}%
\bibitem [{\citenamefont {Fleischhauer}, \citenamefont {Imamoglu},\ and\
  \citenamefont {Marangos}(2005)}]{FleischhauerRevModPhys05}%
  \BibitemOpen
  \bibfield  {author} {\bibinfo {author} {\bibfnamefont {M.}~\bibnamefont
  {Fleischhauer}}, \bibinfo {author} {\bibfnamefont {A.}~\bibnamefont
  {Imamoglu}}, \ and\ \bibinfo {author} {\bibfnamefont {J.~P.}\ \bibnamefont
  {Marangos}},\ }\bibfield  {title} {\enquote {\bibinfo {title}
  {Electromagnetically induced transparency: Optics in coherent media},}\
  }\href {\doibase 10.1103/RevModPhys.77.633} {\bibfield  {journal} {\bibinfo
  {journal} {Rev. Mod. Phys.}\ }\textbf {\bibinfo {volume} {77}},\ \bibinfo
  {pages} {633--673} (\bibinfo {year} {2005})}\BibitemShut {NoStop}%
\bibitem [{\citenamefont {Finkelstein}\ \emph {et~al.}(2023)\citenamefont
  {Finkelstein}, \citenamefont {Bali}, \citenamefont {Firstenberg},\ and\
  \citenamefont {Novikova}}]{FinkelsteinNJP2023}%
  \BibitemOpen
  \bibfield  {author} {\bibinfo {author} {\bibfnamefont {R.}~\bibnamefont
  {Finkelstein}}, \bibinfo {author} {\bibfnamefont {S.}~\bibnamefont {Bali}},
  \bibinfo {author} {\bibfnamefont {O.}~\bibnamefont {Firstenberg}}, \ and\
  \bibinfo {author} {\bibfnamefont {I.}~\bibnamefont {Novikova}},\ }\bibfield
  {title} {\enquote {\bibinfo {title} {A practical guide to electromagnetically
  induced transparency in atomic vapor},}\ }\href
  {http://iopscience.iop.org/article/10.1088/1367-2630/acbc40} {\bibfield
  {journal} {\bibinfo  {journal} {New Journal of Physics}\ } (\bibinfo {year}
  {2023})}\BibitemShut {NoStop}%
\bibitem [{\citenamefont {Grangier}\ \emph {et~al.}(1987)\citenamefont
  {Grangier}, \citenamefont {Slusher}, \citenamefont {Yurke},\ and\
  \citenamefont {LaPorta}}]{GrangierPRL1987SqRotation}%
  \BibitemOpen
  \bibfield  {author} {\bibinfo {author} {\bibfnamefont {P.}~\bibnamefont
  {Grangier}}, \bibinfo {author} {\bibfnamefont {R.~E.}\ \bibnamefont
  {Slusher}}, \bibinfo {author} {\bibfnamefont {B.}~\bibnamefont {Yurke}}, \
  and\ \bibinfo {author} {\bibfnamefont {A.}~\bibnamefont {LaPorta}},\
  }\bibfield  {title} {\enquote {\bibinfo {title} {Squeezed-light--enhanced
  polarization interferometer},}\ }\href {\doibase 10.1103/PhysRevLett.59.2153}
  {\bibfield  {journal} {\bibinfo  {journal} {prl}\ }\textbf {\bibinfo {volume}
  {59}},\ \bibinfo {pages} {2153--2156} (\bibinfo {year} {1987})}\BibitemShut
  {NoStop}%
\bibitem [{\citenamefont {Reim}\ \emph {et~al.}(2011)\citenamefont {Reim},
  \citenamefont {Michelberger}, \citenamefont {Lee}, \citenamefont {Nunn},
  \citenamefont {Langford},\ and\ \citenamefont {Walmsley}}]{ReimPRL2011}%
  \BibitemOpen
  \bibfield  {author} {\bibinfo {author} {\bibfnamefont {K.~F.}\ \bibnamefont
  {Reim}}, \bibinfo {author} {\bibfnamefont {P.}~\bibnamefont {Michelberger}},
  \bibinfo {author} {\bibfnamefont {K.~C.}\ \bibnamefont {Lee}}, \bibinfo
  {author} {\bibfnamefont {J.}~\bibnamefont {Nunn}}, \bibinfo {author}
  {\bibfnamefont {N.~K.}\ \bibnamefont {Langford}}, \ and\ \bibinfo {author}
  {\bibfnamefont {I.~A.}\ \bibnamefont {Walmsley}},\ }\bibfield  {title}
  {\enquote {\bibinfo {title} {Single-photon-level quantum memory at room
  temperature},}\ }\href {\doibase 10.1103/PhysRevLett.107.053603} {\bibfield
  {journal} {\bibinfo  {journal} {Phys. Rev. Lett.}\ }\textbf {\bibinfo
  {volume} {107}},\ \bibinfo {pages} {053603} (\bibinfo {year}
  {2011})}\BibitemShut {NoStop}%
\end{thebibliography}%

\end{document}


\title{Electron Beam Characterization via Quantum Coherent Optical Magnetometry}

\author{Nicolas DeStefano}
\email{ncdestefano@wm.edu}
\author{Saeed Pegahan}
\affiliation{Dept. of Physics, William \& Mary, Williamsburg, Virginia 23187, USA}
\author{Aneesh Ramaswamy}
\affiliation{Stevens Institute of Technology, Hoboken, New Jersey 07030, USA}
\author{Seth Aubin}
\author{T. Averett}
\affiliation{Dept. of Physics, William \& Mary, Williamsburg, Virginia 23187, USA}
\author{Alexandre Camsonne}
\affiliation{Thomas Jefferson National Accelerator Facility, Newport News, Virginia 23606, USA}
\author{Svetlana Malinovskaya}
\affiliation{Stevens Institute of Technology, Hoboken, New Jersey 07030, USA}
\author{Eugeniy E. Mikhailov}
\affiliation{Dept. of Physics, William \& Mary, Williamsburg, Virginia 23187, USA}
\author{Gunn Park}
\author{Shukui Zhang}
\affiliation{Thomas Jefferson National Accelerator Facility, Newport News, Virginia 23606, USA}
\author{Irina Novikova}
\affiliation{Dept. of Physics, William \& Mary, Williamsburg, Virginia 23187, USA}

\date{\today}

\maketitle 

\renewcommand{\thefigure}{S\arabic{figure}}
\renewcommand{\theequation}{S\arabic{equation}}
\setcounter{figure}{0}
\setcounter{equation}{0}
\section*{Supplementary Materials}
\subsection{Nonlinear magneto-optical polarization rotation in Rb vapor} \label{sec:EITcalc}

\begin{figure}[h!]
    \centering
    \includegraphics[width = 0.6\textwidth]{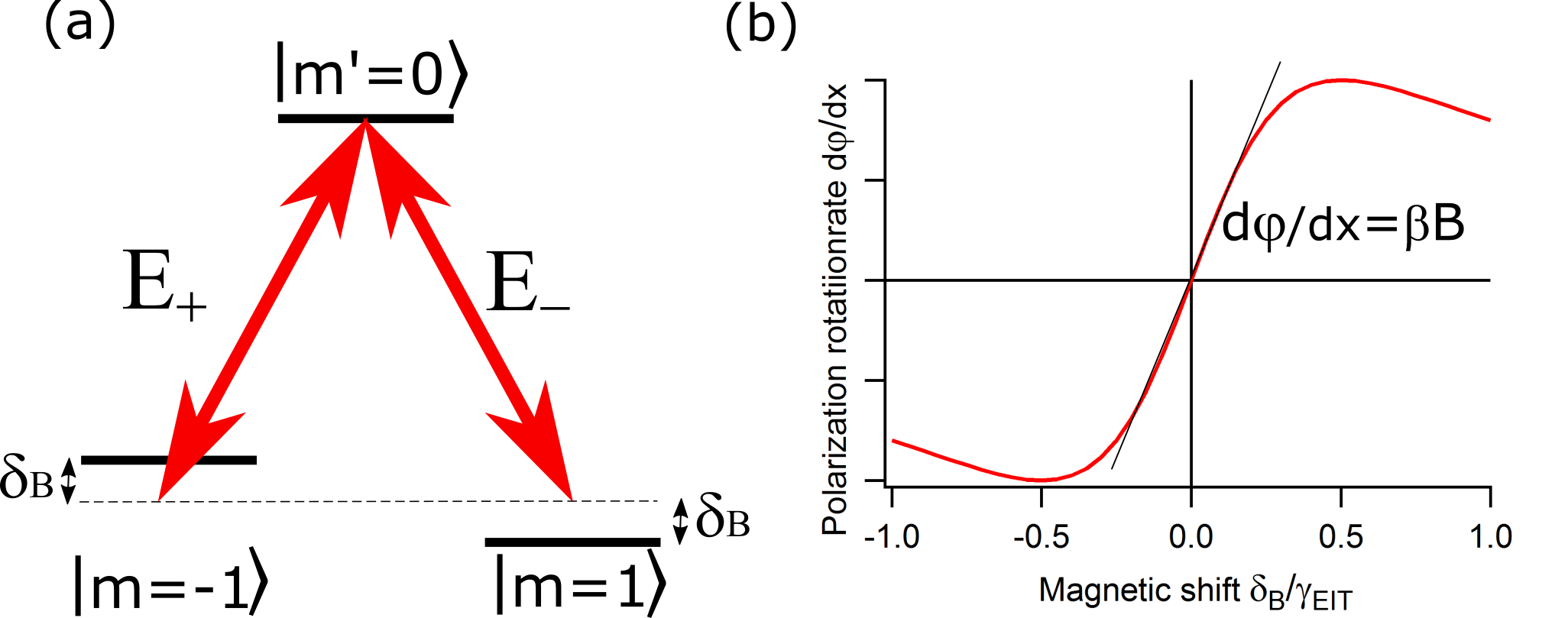}
    \caption{(a) Simplified energy level diagram of the state shift during nonlinear magneto-optical rotation. (b) Typical dependence of the polarization rotation rate $d\varphi/dx$ on the magnetic shift $\delta_B = \gamma B$. Note that near $B=0$ the rotation rate is proportional to the applied field $d\varphi/dx = \beta B$.}
    \label{fig:lambda_system}
\end{figure}

Since accurate theoretical description of nonlinear magneto-optical polarization rotation (NMOR) is well-developed~\cite{Budker_RMP2002,MatskoPRA2003}, here we only summarize its main features relevant to the presented experiment. As the name suggests, the polarization rotation for linearly polarized light arises from the circular birefringence induced in atomic vapor in the presence of the magnetic field. The word ``nonlinear'' in NMOR indicates that the birefringence originates from nonlinear (laser power-dependent) optical susceptibility. Fig.~\ref{fig:lambda_system} shows a simplified NMOR interaction scheme, consisting of two Zeeman sublevels of the ground $5S_{1/2}$ electronic states $|m=\pm1\rangle$, coupled to the $|m'=0\rangle$ excited state $5P_{3/2}$ via two circular components of the linearly polarized near-resonant laser field, forming a so-called $\Lambda$ scheme. While Rb and other alkali metals have richer Zeeman structure, it is well approximated by the simplified three-level system~\cite{MatskoPRA2003}.  

The most remarkable property of such a $\Lambda$ interaction scheme is the existence of a so-called ``dark state'' $|D\rangle = (|m=1\rangle-|m=-1\rangle)/\sqrt{2}$ that can be completely decoupled from the excited state in the absence of the magnetic field (and thus does not produce fluorescence, giving the state its name). Atoms in such state (assuming no decoherence) do not absorb light, giving rise to the effect of electromagnetically induced transparency (EIT)~\cite{harris'97pt,FleischhauerRevModPhys05,FinkelsteinNJP2023}. The resulting transmission resonance, however, is very narrow, so that even small variations of the ground  state energies due to the Zeeman shifts $\delta_B=\gamma B$ produce noticeable drops in transmission and opposite variation in the refractive index for the two circular laser polarization components $n_\pm$. As a result, after propagation through the atomic ensemble of length $L$, the two fields develop opposite phases $\phi_\pm = 2\pi n_\pm L/\lambda$ (here $\lambda$ is the laser wavelength in vacuum), that corresponds to the rotation of the original linear polarization by the angle $\varphi = (\phi_+-\phi_-)/2$. 

One can obtain the accurate solution for the optical susceptibility for the two circular components by solving Maxwell-Bloch equations for the density matrix elements~\cite{Budker_RMP2002,MatskoPRA2003}:

\begin{equation}
    \chi_\pm = \frac{i\alpha_0\lambda}{\pi}\frac{(\Gamma_0\Gamma_{EIT}+4\delta_B^2) - i\delta_B\Gamma_{EIT}}{4\delta_B^2 + \Gamma_{EIT}^2},
\end{equation}

where $\alpha_0=\frac{\mu^2N}{2\epsilon_0\hbar\Gamma}$ is the unsaturated optical resonant absorption, $\mu$ is the dipole moment of the optical transition, $N$ is the atomic density, $\Gamma$ and $\Gamma_0$ are decoherence rates of the excited and ground states, correspondingly, $\Gamma_{EIT}  = \Gamma_0 + \frac{\mu^2}{c\epsilon_0\hbar^2\Gamma}I$ is the power-broadened width of the EIT resonance, where $I$ is the laser intensity. Using this expression, we can determine the variation in the laser intensity and polarization rotation angle as light propagates through the atomic ensemble for small magnetic shift $\delta_B \ll \Gamma_{EIT}$:

\begin{eqnarray}
    \frac{dI}{dx} &=& -\alpha_0\frac{\Gamma_0}{\Gamma_{EIT}},\\ \label{eq:eitabsrate}
     \frac{d\varphi}{dx} &=& \alpha_0\frac{\Gamma \delta_B}{\Gamma_{EIT}^2}.\label{eq:eitrotrate}
\end{eqnarray}

It is easy to see that in the limit of negligible spin decoherence ($\Gamma_0\ll\Gamma_{EIT},\Gamma$) the laser intensity variation is small, and the polarization rotation rate reduces to Eq. 1. 

\subsection{Ultimate beam current measurement sensitivity analysis}

Here we estimate the optimal performance of the proposed sensor. The current performance is severely affected by the technical noises in the optical detection.
As the dark noise of the CCD camera, used here, is higher than for a balanced photo-diode detector (BPD), we consider the latter to estimate the best achievable sensitivity. The smallest measurable $e$-beam current ($\delta I_e$) is estimated to correspond to the signal to noise ratio equal to one. In our case, it is determined by the smallest measurable polarization rotation angle ($\delta \varphi$) and by the response of our system $S$, i.e. the polarization rotation angle per current carried by the $e$-beam:

\begin{equation}
	\label{eq:current_sensitivity}
	\delta I_e = \left(\frac{d \varphi}{d I_e}\right)^{-1} \delta \varphi = \frac{1}{S} \delta \varphi
\end{equation}

To estimate the response of the system, we varied the emission current of our electron gun in the range of 0 - 90~$\mu$A. We observe the linear relationship between the measured rotation angle and the emission current with a slope $S=0.21$~mrad/$\mu$A. (we note that the emission current is consistently 50\% larger than the current indicated by the Faraday cup due to the losses along the electron beam propagation path, such as beam line obstacles, back-scattered electrons, etc.). 

\begin{figure}[t]
    \centering
    \includegraphics[width = 0.55\textwidth]{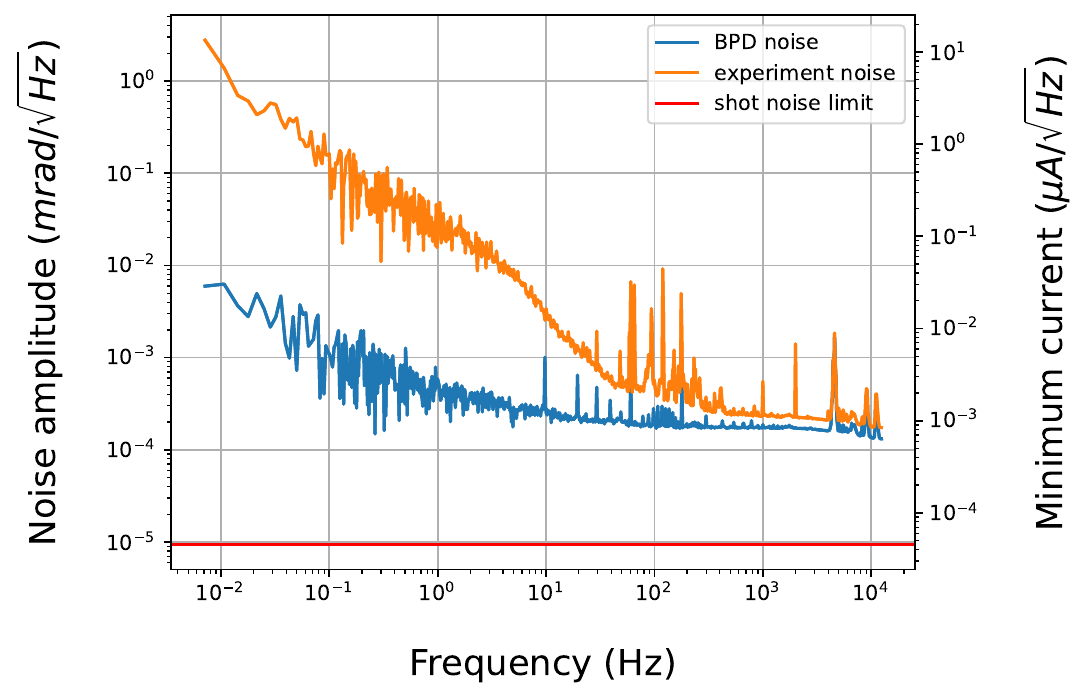}
    \caption{Fourier transformed signals from the BPD in the absence of polarization rotation. Plotted are the cases of typical experimental conditions (experiment noise) compared to the dark noise of the BPD (BPD noise). The red line in the bottom shows the shot-noise limited sensitivity. The right axis defines the minimum detectable current, based on the minimum detectable rotation angle of the left axis.}
    \label{fig:fft_trace}
\end{figure}

The smallest measured rotation angle (per unit of bandwidth) is fundamentally set by the shot noise of the laser beam, and depends on the laser power $P$~\cite{GrangierPRL1987SqRotation} as

\begin{equation}
	\label{eq:light_shot_noise}
	\delta \varphi = \sqrt{ \frac{ 2 h c/\lambda } { \eta P } }= \sqrt{ \frac{ 2 e } { I_{ph} } }
\end{equation}

where $h$ is the Planck constant; $c$ is the speed of light; $\lambda$ is the laser probe wavelength, $\eta$ is the quantum efficiency of the detector, $e$ is charge of the electron and $I_{ph}$ is total photo-current generated in the BPD. It is easy to see that the expression under the square root is just the number of photo-electrons generated by light in the photodiode, per unit of time.

Combining equations~\ref{eq:current_sensitivity} and~\ref{eq:light_shot_noise}, we calculated shot noise-limited sensitivity for our setup to be 400~pA$/\sqrt{\text{Hz}}$ (see Fig.~\ref{fig:fft_trace}).  In practice, the sensitivity limit is actually set by the classical noise sources: residual light intensity noise, mechanical vibrations, flicker noise in electronics, that make it hard to reach the shot noise limited performance. In our case, the performance of the BPD was limited by the photodiode dark noise at high frequencies to about 1~nA$/\sqrt{\text{Hz}}$ (see Fig.~\ref{fig:fft_trace}). At lower frequencies, we see performance degradation with a characteristic $1/f$ noise.

There are a few straightforward steps for improving the experimental sensitivity. From Fig. \ref{fig:fft_trace} it is clear that increasing the detection frequency can reduce the noise dramatically. In the current experiment, the data was recorded at 1~Hz rate due to the limited speed of the electron beam modulator, but in practice many charged particle beams are pulsed at a much faster rate. The current detection scheme limits the speed of the atomic response by the excited state lifetime ($\approx 30$~ns); however, using a different, off-resonant two-photon Raman scheme, one should be able to achieve sub-ns response~\cite{ReimPRL2011}. Moreover, some laser-related noises can be suppressed by monitoring a reference probe beam unaffected by the charged particles. Subtracting such a reference from the signal laser beam can help to eliminate the background signals not pertaining to the electron beam.

\subsection{Rubidium Fluorescence Due To Electron Scattering}

\begin{figure}[h!]
    \centering
    \includegraphics[width=0.85\textwidth]{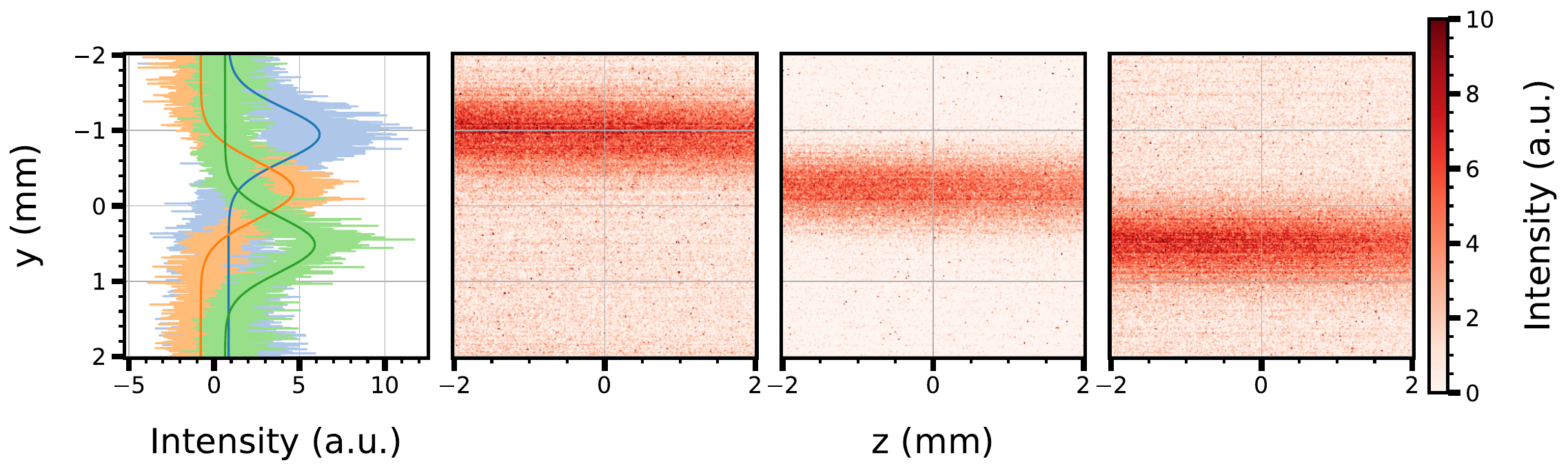}
    \caption{Rubidium fluorescence induced by the electron beam at three vertical positions. Note that the vertical displacement here is not equivalent to the e-beam positions imaged in Fig. 2.}
    \label{fig:fluorescence_imaging_results}
\end{figure}

To verify the electron beam profiles, we use fluorescence imaging to observe the scattering of high-energy electrons from rubidium atoms. When an incoming electron approaches an atom, there is a small probability for ionization. As ionized Rb atoms return to their original state, they emit photons that can be detected by a camera. By monitoring this fluorescence using an auxiliary imaging system with a magnification $1$, we obtained independent verification of the $e$-beam characteristics. Fig. \ref{fig:fluorescence_imaging_results} shows examples of acquired fluorescence images due to the $e$-beam for different vertical positions. We employed a camera with a 30-second exposure time to capture the fluorescence images, considering the low number of excited atoms that emit photons through fluorescence. Moreover, the quantum efficiency (QE) of the camera at 780 nm, which is approximately $25\%$, was taken into account. 

\bibliography{bibliography}